\documentclass[fleqn,usenatbib]{mnras}

\usepackage{newtxtext,newtxmath}

\usepackage[T1]{fontenc}

\DeclareRobustCommand{\VAN}[3]{#2}
\let\VANthebibliography\thebibliography
\def\thebibliography{\DeclareRobustCommand{\VAN}[3]{##3}\VANthebibliography}

\usepackage{graphicx}	
\usepackage{amsmath}	
\usepackage{threeparttable}
\usepackage{CJK}

\title[Magnetic Fields of G11]{The Role of Magnetic Fields in the Formation of the Filamentary Infrared Dark Cloud G11.11--0.12}

\author[Chen, Zhiwei et al.]{
Zhiwei Chen (陈志维),$^1$\thanks{E-mail: zwchen@pmo.ac.cn}
Ramotholo Sefako,$^2$
Yang Yang,$^{3}$
Zhibo Jiang,$^{1}$
Yang Su,$^1$
Shaobo Zhang $^1$  
\newauthor
and Xin Zhou$^1$
\\
$^{1}$Purple Mountain Observatory, Chinese Academy of Sciences, 10 Yuanhua Road, 210023 Nanjing, China\\
$^{2}$South African Astronomical Observatory, PO Box 9, Observatory 7935, Cape Town, South Africa\\
$^{3}$Center for Astrophysics, Guangzhou University, Guangzhou 510006, China
}

\date{Accepted 2023 July 21. Received 2023 July 20; in original form 2022 July 06}

\pubyear{2023}

\begin{document}
\begin{CJK*}{UTF8}{gbsn}
\label{firstpage}
\pagerange{\pageref{firstpage}--\pageref{lastpage}}
\maketitle

\begin{abstract}
We report on the near-infrared polarimetric observations of G11.11--0.12 (hereafter G11) obtained with SIRPOL on the 1.4 m IRSF telescope. The starlight polarisation of the background stars reveals the on-sky component of magnetic fields in G11, and these are consistent with the field orientation observed from polarised dust emission at $850\,\mu$m. The magnetic fields in G11 are perpendicular to the filament, and are independent of the filament's orientation relative to the Galactic plane. The field strength in the envelope of G11 is in the range $50-100\,\mu$G, derived from two methods. The analyses of the magnetic fields and gas velocity dispersion indicate that the envelope of G11 is supersonic but sub-Alfv{\'e}nic. The critical mass-to-flux ratio in the envelope of G11 is close to 1 and increases to $\gtrsim 1$ on the spine of G11. The relative weights on the importance of magnetic fields, turbulence and gravity indicate that gravity dominates the dynamical state of G11, but with significant contribution from magnetic fields. The field strength, $|\mathbf{B}|$, increases slower than the gas density, $n$, from the envelope to the spine of G11, characterized by $|\mathbf{B}|\propto n^{0.3}$. The observed strength and orientation of magnetic fields in G11 imply that supersonic and sub-Alfv{\'e}nic gas flow is channelled by the strong magnetic fields and is assembled into filaments perpendicular to the magnetic fields. The formation of low-mass stars is enhanced in the filaments with high column density, in agreement with the excess of low-mass protostars detected in the densest regions of G11.
\end{abstract}



\section{Introduction}           
\label{sect:intro}
Filaments are ubiquitous interstellar medium (ISM) structures in galaxies. Their formation is, however, still a controversial topic \citep[see][for reviews]{2014prpl.conf...27A,2022arXiv220309562H}. Gas compression due to supersonic turbulence and/or colliding gas flow is supposed to be responsible for the formation of filaments \citep[e.g.][]{1994ApJ...423..681V,2001ApJ...553..227P,2006ApJ...643..245V,2006ApJ...648.1052H}. Once a compressed layer more massive than the average Jeans mass is formed, gravitational contraction starts to take control of the evolution, and naturally produce filaments in the cloud and clumps inside the filaments \citep[e.g.][]{2014ApJ...791..124G,2022MNRAS.512.4715N}. The clumps in filaments are potential sites of star formation triggered by gravitational collapse \citep[e.g.][]{2019MNRAS.485.4686W,2020RAA....20..115Y,2021ApJ...922..144Y}. Star formation activity such as in young stellar objects are mostly found in filaments \citep[e.g.][]{2012A&A...547A..49R,2019A&A...622A..52Z}. In addition to turbulence and gravity, magnetic field also plays an important role in the formation and evolution of filaments \citep[e.g.][]{1993ApJ...404L..83H,1996ApJ...472..673G}. In the past two decades, magnetohydrodynamic (MHD) simulations and observations of magnetic fields in filaments have shown that magnetic fields play an important role in shaping filaments and regulating the subsequent star formation inside them \citep[e.g.][]{2008ApJ...680..428G,2008ApJ...687..303I,2009ApJ...695..248H,2011ApJ...734...63S,2011ApJ...741...21C,2013A&A...556A.153H,2013ApJ...774..128S,2015MNRAS.452.2410S,2015MNRAS.451.3340K,2016ApJ...832..186S,2018MNRAS.474.4824Z,2018MNRAS.480.3916M,2019MNRAS.485.4509L,2020MNRAS.499.4785K,2021A&A...647A..78A,2022ApJ...925..196I,2022MNRAS.510.6085L}.

The orientation of magnetic fields with respect to filaments shows a bimodal distribution, either parallel or perpendicular \citep[e.g.][]{2013MNRAS.436.3707L,2014ApJ...789...82C,2014prpl.conf..101L,2016A&A...586A.135P,2017A&A...603A..64S,2019A&A...629A..96S}. The bimodal orientation between magnetic fields and filaments is an indication of the physical link between magnetic fields and filaments, otherwise orientation between them should be following a random distribution. A flip in relative orientation, from parallel to perpendicular, occurs around a transition column density of $\sim10^{21-22}\,\mathrm{cm}^{-2}$ \citep[e.g.][]{2016ApJ...829...84C,2016MNRAS.460.1934M,2016A&A...586A.138P,2017A&A...603A..64S,2020MNRAS.496.4546H}, or a number density of $\sim10^{3-4}\,\mathrm{cm}^{-3}$ \citep[e.g.][]{2019ApJ...878..110F,2021ApJ...918...39L}. Below the transition column density, filaments or striations are parallel to magnetic fields. Above the transition column density, dense filaments are perpendicular to magnetic fields. \citet{2020NatAs...4.1195P} reported a further transition in relative orientation between magnetic fields and filament, a return to parallel alignment at $A_V>21$ mag (or column density $\gtrsim2\times10^{22}\,\mathrm{cm^{-2}}$) in parts of the Serpens South cloud. This parallel alignment at a higher column density might be caused by gas flow towards the magnetically supercritical parts of filaments driven by gravitational collapse \citep{2018MNRAS.480.2939G,2020NatAs...4.1195P}. \citet{2020MNRAS.497.4196S} suggested that the transition from parallel to perpendicular occurs at a column density corresponding to the critical mass-to-flux ratio ($M/\Phi$) of a cloud close to 1. Magnetically supercritical filaments, or filaments where gravity is stronger than magnetic field, are preferentially perpendicular to magnetic fields. On the other hand, magnetic fields in subcritical filaments channel the gas flow along the field lines, given that the cross field motions are resisted by Lorentz force. 
The bimodal orientation between filaments and magnetic field also holds for sub-Alfv{\'e}nic clouds, in which the ratio of turbulent velocity dispersion $\delta v$ to Alfv{\'e}n speed $v_\mathrm{A}$ is smaller than 1. This ratio is referred as the Alfv{\'e}n Mach number ${\cal M}_\mathrm{A}=\delta v/v_\mathrm{A}$. In clouds with super-Alfv{\'e}nic (${\cal M}_\mathrm{A}>1$) and supersonic (Mach number ${\cal M}>1$) turbulence, filaments may be formed through converging flows in which magnetic fields are compressed; in this case filaments and magnetic fields are parallel and twisted \citep{2021MNRAS.503.5425B}. Strong turbulence could be produced by violent feedback from massive stars (\ion{H}{ii} regions, stellar winds, supernova explosions). The swept-up shells by expanding \ion{H}{ii} regions look like filaments with column density higher than $\sim10^{22}\,\mathrm{cm}^{-2}$. The magnetic fields within the shells are compressed by the supersonic shock waves, and are parallel to the filament-like shells \citep[e.g.][]{2007ApJ...671..518K,2011MNRAS.414.1747A,2017ApJ...838...80C,2022_chen_rcw120}. The magnetic field measurements are rather limited compared to other properties of filaments (e.g. mass, length, velocity dispersion). The poorly understood magnetic fields of filaments thus prevent a more comprehensive understanding of the dynamical states of filaments. Given the potential wealth of information stored in the magnetic field structure and strength on the formation and evolution of filaments, a more detailed investigation of the magnetic field properties in individual filaments is warranted. 

Infrared dark clouds (IRDCs) are a type of dense structures, mostly appearing as filaments. IRDCs are prominent absorption features against the infrared background in the Galactic plane, and are typically located at distances of a few kpc \citep{1998ApJ...508..721C,2006ApJ...639..227S,2009A&A...505..405P,2010A&A...515A..42R}. Massive dust condensations and cores, precursors to massive star and cluster formations, are ubiquitously found within IRDCs \citep{2010A&A...518L..95H,2012A&A...547A..49R}. The evolution of IRDCs is expected to be governed by the combination of gravity, turbulence, and magnetic fields. However, the relative importance of these combined effects is difficult to determine. While gravity and turbulence can be explored from (sub)millimeter observations, detecting magnetic fields of IRDCs is always a challenge. 

G11.11--0.12, hereafter G11, is among the first IRDCs discovered, and with the darkest shadows in the Galactic plane \citep{1998ApJ...508..721C}. It is located at a distance of 3.6 kpc, has a length of $\sim30$ pc, a mass of $\sim10^5\,M_\odot$, and is known to host multiple star-forming sites \citep{2006A&A...447..929P,2010A&A...518L..95H,2011A&A...529A.161G,2014A&A...570A..51S,2014ApJ...796..130R,2014MNRAS.439.3275W,2019A&A...622A..54P,2019ApJ...886..102S,2020ApJ...903..119L,2021ApJ...913..131T}. The density structure of G11 can be fitted by a central massive cylinder with an extended envelope \citep{2003ApJ...588L..37J,2013A&A...557A.120K}. The spine of G11 has a linear mass density of $\sim600 M_\odot\,\mathrm{pc}^{-1}$ that greatly exceeds the critical value of a self-gravitating, non-turbulent cylinder \citep{2013A&A...557A.120K}. The hierarchical fragmentation in G11 indicates the dominant role of gravity that has started gravitational collapse \citep{2013A&A...557A.120K,2014MNRAS.439.3275W,2015A&A...578A..29S,2015A&A...573A.119R,2019A&A...622A..54P}. The column density probability distribution function of G11 has a power-law tail in the high column density end, indicative of a gravity-dominated stage \citep{2015A&A...578A..29S,2017ApJ...840...22L}. Signatures of gravitational collapse and longitudinal gas flow have been reported in G11 \citep{2014A&A...564A.106T,2014A&A...571A..53B,2015A&A...578A..29S}. It is likely that global collapse in filament scale and fragmentation in clump/core scale are simultaneously occurring in G11. Given the above observational properties, G11 is likely at a relatively late stage of filament formation. Magnetic fields, inherited from the natal ISM that G11 condensed out of, may provide additional information about the formation of G11.

\citet{2004ApJ...616..925F} proposed that G11 is radially supported by predominantly poloidal magnetic fields. The magnetic fields towards G11 were first determined using $850\,\mu$m polarisation data \citep{2015ApJ...799...74P}, taken with the polarimeter SCUPOL on the Jame Clerk Maxwell Telescope (JCMT) \citep{2000ApJ...543L.157C,2003MNRAS.340..353G}. The areas observed with JCMT/SCUPOL cover a small part of the spine of G11, which is bright and compact in submillimeter wavelengths (see Figure 1 in \citet{2015ApJ...799...74P}). The magnetic fields towards the densest part of G11 at sub-parsec scale are perpendicular to the long axis of the filament \citep{2015ApJ...799...74P}. Generally, magnetic fields at dense core scales ($\lesssim0.1$ pc) are either aligned with or perpendicular to the parsec scale magnetic fields \citep{2014ApJ...792..116Z,2020MNRAS.494.1971C,2021ApJ...912L..27E}. The link between magnetic fields in the spine of G11 (sub-parsec scale) and across the filament (parsec scale) remains unknown.

The submillimeter wavelength polarisation data from Planck mission \citep{2011A&A...536A...1P} are sensitive to large scale extended structures. However, the angular resolution of Planck 353 GHz polarisation data is worse than $\sim5\arcmin$,  which means that G11 is not well resolved by Planck data. The Planck polarisation data are the integrated contributions from dust emission along the same line of sight (LOS) to G11. Starlight polarisation, especially in the near-IR wavelengths, has the potential to reveal magnetic fields in the diffuse regions of G11. The polarisation degrees measured in multiple wavelengths were found to follow the Serkowski relation \citep{1975ApJ...196..261S,1992ApJ...386..562W},
confirming the interstellar origin for starlight polarisation. When the background starlight polarisation is carefully analysed, the starlight polarisation due to dust extinction in G11 carries the information of dust alignment. In diffuse ISM, elongated dust grains with short axes parallel to magnetic fields can be explained by the radiative torque mechanism \citep{2007MNRAS.378..910L,2015ARA&A..53..501A}. The linear polarisation of starlight due to dust extinction is widely used to probe for the on-sky component of magnetic fields (hereafter $B_\mathrm{sky}$), which is parallel to starlight polarisation angle. In this paper, we present the large scale magnetic fields of G11, derived from near-IR starlight polarisation of background stars. The study of the large scale magnetic fields is essential to explore the effects of magnetic field in the formation and evolution of G11.

\section{Data Acquisition}
\label{sect:Obs}
\subsection{IRSF/SIRPOL Observations}
\label{sirpol}
\begin{figure*} 
   \centering
   \includegraphics[width=0.99\textwidth]{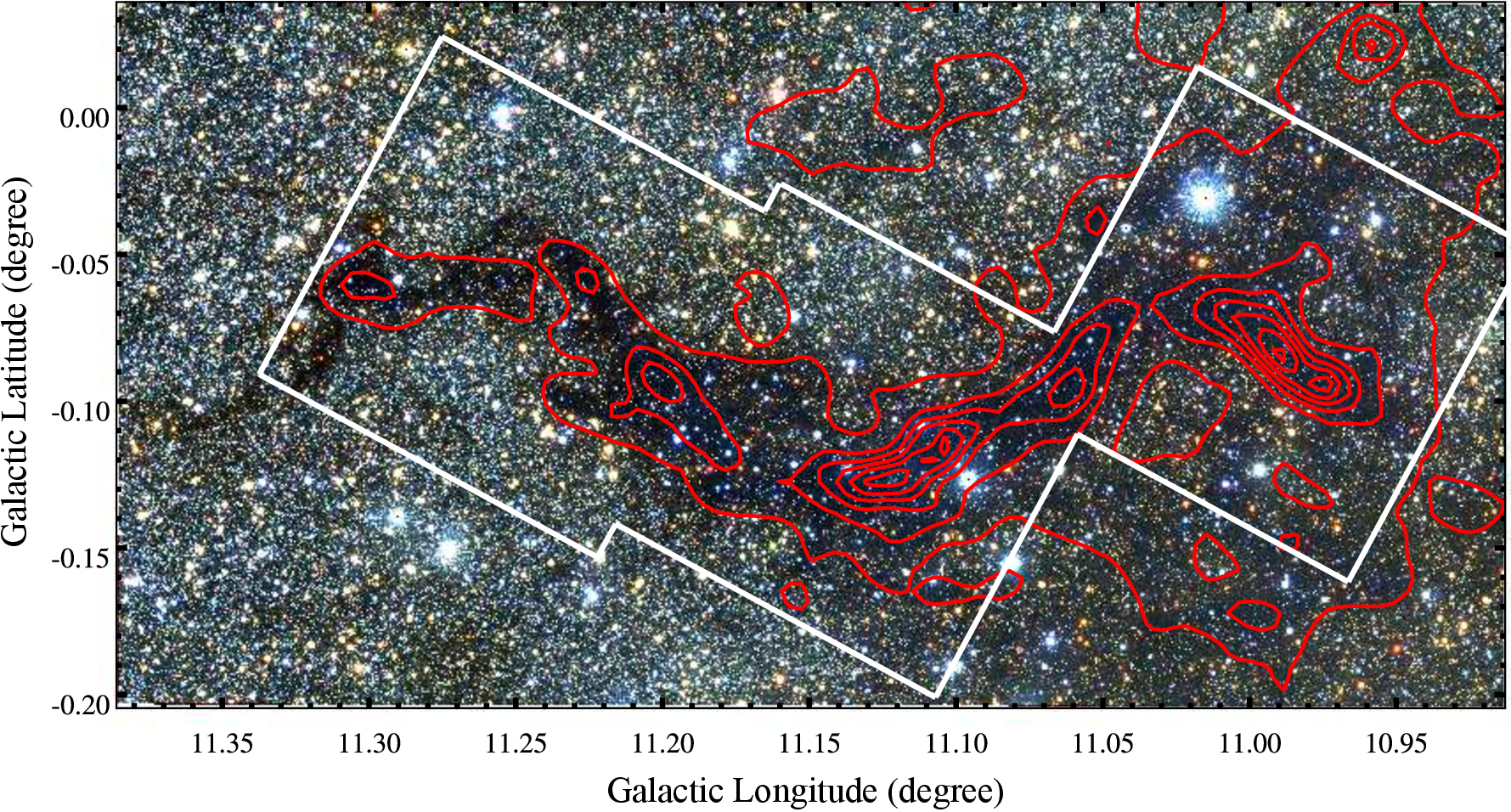}
   \caption{Three-colour composite image of G11, made from the VVVX $J$ (blue), H (green), and $K_s$ (red) images, overlaid with the red contours of $N(\mathrm{H_2})_\mathrm{dust}$ from $2\times10^{22}\,\mathrm{cm^{-2}}$ to $8\times10^{22}\,\mathrm{cm^{-2}}$ in steps of $10^{22}\,\mathrm{cm^{-2}}$. The coverage of the IRSF observations is marked by the inserted grid shape made from three different pointings of IRSF, within the white lines and as discussed in Section \ref{sirpol}. } 
   \label{Fig:VVV_JHK}
 \end{figure*}

The data were taken with SIRPOL on the 1.4 m InfraRed Survey Facility (IRSF) telescope at the South African Astronomical Observatory, Sutherland, South Africa. SIRPOL is a single-beam polarimeter capable of simultaneous imaging polarisation observations in the near-IR $JHK_s$ bands with an $\sim 8\arcmin\times8\arcmin$ field of view (FOV)\citep{2003SPIE.4841..459N,2006SPIE.6269E..51K}. The observations were made during the night of 2018 July 21. To cover the entire extent of G11, we designed a mosaic of three frame positions around the location of G11, as shown in Figure~\ref{Fig:VVV_JHK}. In each frame position, three sets of observations were taken. In each set, the telescope was dithered at 10 positions, each with 30\,s exposure time at 4 wave-plate angles in a sequence of $0\degr$, $45\degr$, $22.5\degr$, and $67.5\degr$. The total integration time was 1200\,s per wave-plate angle in each of the three frame positions. The total observation time in each of the three frame positions was 83 min, taking observation overheads into account. The total observation time for all the three frame positions was 4.1 hr. The airmass during the observations was in the range of $1.03-1.60$. Observations with lower airmass generally have narrower full width at half maximum (FWHM). The average FWHM for the point sources in all the three frame positions were determined be ranging from around 3 to 4 pixels, or $\sim1\farcs5$--$1\farcs8$ ($0\farcs45$ per pixel).

In the same night of the IRSF observations, a field from the Galactic Plane Infrared polarisation Survey (GPIPS)  \citep{2012ApJS..200...19C} was observed with SIRPOL just before the observations for G11. The original field GP0006 from GPIPS is centered on $R.A.(J2000)=18^\mathrm{h}26^\mathrm{m}11\fs2$, $Decl.(J2000)= -13\degr19\arcmin 57\arcsec$, and has a $10\arcmin\times10\arcmin$ FOV. The SIRPOL images for the GP0006 field are centered on the inner $\sim 8\arcmin\times8\arcmin$ region, slightly smaller than the original FOV of the GP0006 field. 

The data reduction and aperture photometry for the $JHK_s$ polarimetric observations of G11 and the GP0006 field are referred to in \citet{2022_chen_rcw120}. Following the procedures used in determining polarisation measurements in \citet{2022_chen_rcw120}, we obtained the $JHK_s$ polarisation measurements of sources towards G11 and the GP0006 field. The polarisation measurements have a positive bias because the polarisation degree derived from $\sqrt{Q^2+U^2}$ is always positive, even though the normalized Stokes parameters $Q$ and $U$ can be either positive or negative. Following the methods of correcting this positive bias \citep{1974ApJ...194..249W,2015JAI.....450005H}, the bias-corrected polarisation is $P=\sqrt{Q^2+U^2 - \delta P^2}$. In the following texts, the bias-corrected polarisation is labeled as $P$, unless otherwise noted. The polarisation position angle $\theta_\mathrm{PA}$ is measured counterclockwise from the north to the east using equatorial coordinates. The $1\sigma$ uncertainty of $\theta_{PA}$ depends on the polarisation signal-to-noise ratio $\frac{P}{\delta P}$. For simplicity, the $1\sigma$ uncertainty of $\theta_{PA}$ is obtained from the standard formula  $\delta \theta_{PA}=28\fdg6\,\delta P/P$. For sources with $\frac{P}{\delta P}\geqslant6$, the $1\sigma$ uncertainty of $\theta_{PA}$ obtained as outlined above is reliable \citep{1993A&A...274..968N}. For lower SNR ($2\leqslant\frac{P}{\delta P}<6$), the $1\sigma$ uncertainty of $\theta_{PA}$ is underestimated, with smaller deviation for higher $\frac{P}{\delta P}$. More than 85\% of the observed sources towards G11 in all the $JHK_s$ bands meet the criterion $\frac{P}{\delta P}\geqslant3$, which means that majority of the sources considered in this work have $1\sigma$ uncertainty of $\theta_{PA}$ of less than $10\degr$.

The $H$-band starlight polarisation obtained with SIRPOL towards the GP0006 field of GPIPS are compared to the GPIPS catalog \citep{2012ApJS..200...19C,2020ApJS..249...23C}.  We found 13 polarised stellar sources common in both our SIRPOL observations and the GPIPS catalog. These stellar sources satisfy $P_H/\delta P_H \geqslant6$ and 2MASS $J-H$ $\geqslant0.9$, and have $P_H$ in the range from $1\%$ to $9\%$ and $\theta_\mathrm{PA}$ in the range from $10\degr$ to $180\degr$. For the 13 stellar sources, the mean difference in $P_H$ relative to the GPIPS value is $-6.2\%$ with a standard deviation of 16.3\%; the mean difference in $\theta_\mathrm{PA}$ relative to the GPIPS value is $7\fdg8$ with a standard deviation of $7\fdg6$. The SIRPOL observations of the GP0006 field show tight consistency with the GPIPS catalog. For the simultaneous observations in the $JHK_s$ bands with the IRSF telescope, the $JK_s$-band starlight polarisation are aligned in polarisation orientations with the $H$-band polarisation. For tracing the on-sky magnetic field component, $B_\mathrm{sky}$, only the polarisation orientations are utilized. The starlight polarisation in the $JHK_s$ bands of the \ion{H}{ii} regions M17 and RCW 120 show consistency in polarisation orientations in all three bands \citep{2012PASJ...64..110C,2022_chen_rcw120}, though a dependence of polarisation degree with the wavelength was observed \citep{2012PASJ...64..110C}.

\subsection{VVVX photometry}
We used the near-IR photometry from the extension of the Vista Variables in the V{\'i}a L{\'a}ctea (hereafter VVVX) survey \citep{2010NewA...15..433M}. G11 was covered in tile e0954 as part of the VVVX Data Release 1. The $JHK_s$ images and catalogs of the e0954 tile were retrieved from the European Southern Observatory (ESO) Science Archive Facility \footnote{The VVVX datasets are available at https://doi.eso.org/10.18727/archive/68}. The VVVX $JHK_s$ catalogs include magnitudes from a series of aperture diameters, ranging from $1\arcsec$ to $24\arcsec$ (aperture 1 to aperture 13). The aperture 3 corresponds to an aperture diameter of $2\arcsec$. We utilized the aperture 3 measurements of the VVVX photometric catalog. The calibrated $JHK_s$ magnitudes can be derived from the aperture 3 fluxes using the following equation:
\begin{align}
CalMag = MAGZPT - 2.5\log{Flux} - APCORN,
\end{align}
where the zero point calibration $MAGZPT$ and aperture correction term $APCORN$ of aperture 3 are integrated in the same VVVX catalog \footnote{See https://www.eso.org/rm/api/v1/public/releaseDescriptions/187}. 

\subsection{CO Molecular Line Data of G11}\label{Sect:CO}
The CO and CO isotopologues $J$=1-0 data of G11 were obtained from the Milky Way Imaging Scroll Painting (MWISP) project, an unbiased Galactic plane CO survey in the northern sky using the 13.7 m millimeter-wavelength telescope at Delingha, China \citep[see][for more details]{2019ApJS..240....9S}. A linear baseline was subtracted from the $^{12}$CO, $^{13}$CO, and C$^{18}$O $J$=1-0 spectra. After the calibration of the main beam efficiency of $\sim50\%$, the reduced position-position-velocity (PPV) cubes with a $30\arcsec\times30\arcsec$ grid are on the intensity scale of main beam temperature, $T_\mathrm{MB}$. The measured root-mean-square (rms) noise is $\sim 0.5\,\mathrm{K}$ for $^{12}$CO $J$=1-0 spectra and $\sim0.3\,\mathrm{K}$ for $^{13}$CO/C$^{18}$O $J$=1-0 spectra at a channel width of $61\,\mathrm{kHz}$ ($\approx 0.16-0.17\,\mathrm{km s^{-1}}$). The angular resolutions of the CO PPV cubes are close to $50\arcsec$. 

Similar to the velocity range of $26 - 35 \,\mathrm{km\,s^{-1}}$ of $^{12}$CO $J$=3-2 emission presented in \citet{2015A&A...578A..29S}, we utilized a wider velocity range of $25-37\,\mathrm{km\,s^{-1}}$, based on the $^{12}$CO and $^{13}$CO $J$=1-0 emission. With this velocity range, we derived the moment 0 (integrated intensity), moment 1 (velocity), and moment 2 (line width) maps of the $^{12}$CO, $^{13}$CO and C$^{18}$O $J$=1-0 emissions from the MWISP CO PPV cubes. The excitation temperature, $T_\mathrm{ex}$, of the CO molecular gas was derived from the $^{12}$CO $J$=1-0 spectra, following Equation 9 in \citet{2022_chen_rcw120}. The pixel-wise $T_\mathrm{ex}$ map across G11 is uniform with a median of $12\,\mathrm{K}$, and a standard deviation of $1\,\mathrm{K}$. Assuming that the molecular gas is in the state of local thermal equilibrium (LTE), the $^{13}$CO and C$^{18}$O gas share the same $T_\mathrm{ex}$ of the $^{12}$CO gas. In the state of LTE, we derived the column densities of $^{13}$CO and C$^{18}$O gas ($N(\mathrm{^{13}CO})$ and $N(\mathrm{C^{18}O})$) according to Equation 10 in \citet{2022_chen_rcw120}. Following Equation 11 in \citet{2022_chen_rcw120}, the optical depth $\tau_\mathrm{13}$ of the $^{13}$CO emission reaches a maximum value $\approx 1$ towards the densest regions of G11. In the remaining parts of lower density in G11, $\tau_\mathrm{13}$ varies mostly between 0.3 and 0.7. The $N(\mathrm{^{13}CO})$ was then multiplied by the optical depth correction factor, $\frac{\tau_{13}}{1-\exp{(-\tau_{13})}}$. We simply assumed that the optical depth of C$^{18}$O $J$=1-0 emission is $\ll 1$. Thus the optical depth correction factor for the C$^{18}$O emission is $\approx1$. Assuming the H$_2$ to $^{13}$CO and C$^{18}$O abundance ratios as $5\times10^5$ and $3.5\times10^6$, respectively \citep[][and references therein]{2022_chen_rcw120}, we derived molecular hydrogen column density $N(\mathrm{H_2})$ from $N(\mathrm{^{13}CO})$ and $N(\mathrm{C^{18}O})$ respectively. The moment 2 map of the $^{13}$CO emission evaluates the velocity dispersion $\delta v$ across G11, which has a narrow distribution with a median of $1.9\,\mathrm{km\,s^{-1}}$ and a standard deviation of $0.2\,\mathrm{km\,s^{-1}}$. Meanwhile the sound speed $c_s$ is around $0.2\,\mathrm{km\,s^{-1}}$ across G11, corresponding to a rather uniform $T_\mathrm{ex}\sim12\,\mathrm{K}$ derived from the $^{12}$CO line emission. The molecular gas is supersonic, characterized by the Mach number ($\mathcal{M}=\delta v/c_s$) $>6$ across G11. The column density, velocity field and velocity dispersion of molecular gas derived from CO PPV cubes and magnetic fields traced by near-IR starlight polarisation are the fundamental parameters characterizing the dynamical state of G11. 

\subsection{Column Density Map from Herschel}
The $N(\mathrm{H_2})_\mathrm{dust}$ map of G11 was constructed from the modified blackbody fit to the Herschel maps at 160, 250, 350 and 500 $\mu$m \citep[see][for further details]{2015A&A...578A..29S}. The $N(\mathrm{H_2})_\mathrm{dust}$ map of G11 was smoothed to an angular resolution of $\sim36\arcsec$, slightly better than the angular resolution of the MWISP CO data.

\section{Results}\label{Sect:res}
Figure~\ref{Fig:VVV_JHK} shows a three-colour composite image of G11 from the VVVX observations. The combined IRSF observations cover the entire G11 filamentary dark cloud  
and sufficient areas around the filament, an indication that  the near-IR polarisation observations span the entire dense filament and surrounding fields of lower gas emission (see also Figure~\ref{Fig:NH2_KPOL}). The red contours of $N(\mathrm{H_2})_\mathrm{dust}$, superimposed on the composite image in Figure~\ref{Fig:VVV_JHK}, also cover the entire filamentary region. 

The near-IR polarisation observations from the IRSF telescope can potentially be used to trace magnetic fields in the surrounding fields of lower density. However, the dense filament in G11.11-0.12 has $N(\mathrm{H_2})_\mathrm{dust}$ well above $4\times10^{22}\,\mathrm{cm^{-2}}$, which is beyond the starlight polarisation detection limit of the IRSF (a 1.4 m class telescope) in the near-IR wavelengths.   Near-IR polarisation observations by a larger telescope may have sufficient sensitivity to detect the faint polarised starlight of background stars lying behind the spine of G11.

\begin{figure*} 
   \centering
   \includegraphics[width=0.99\textwidth]{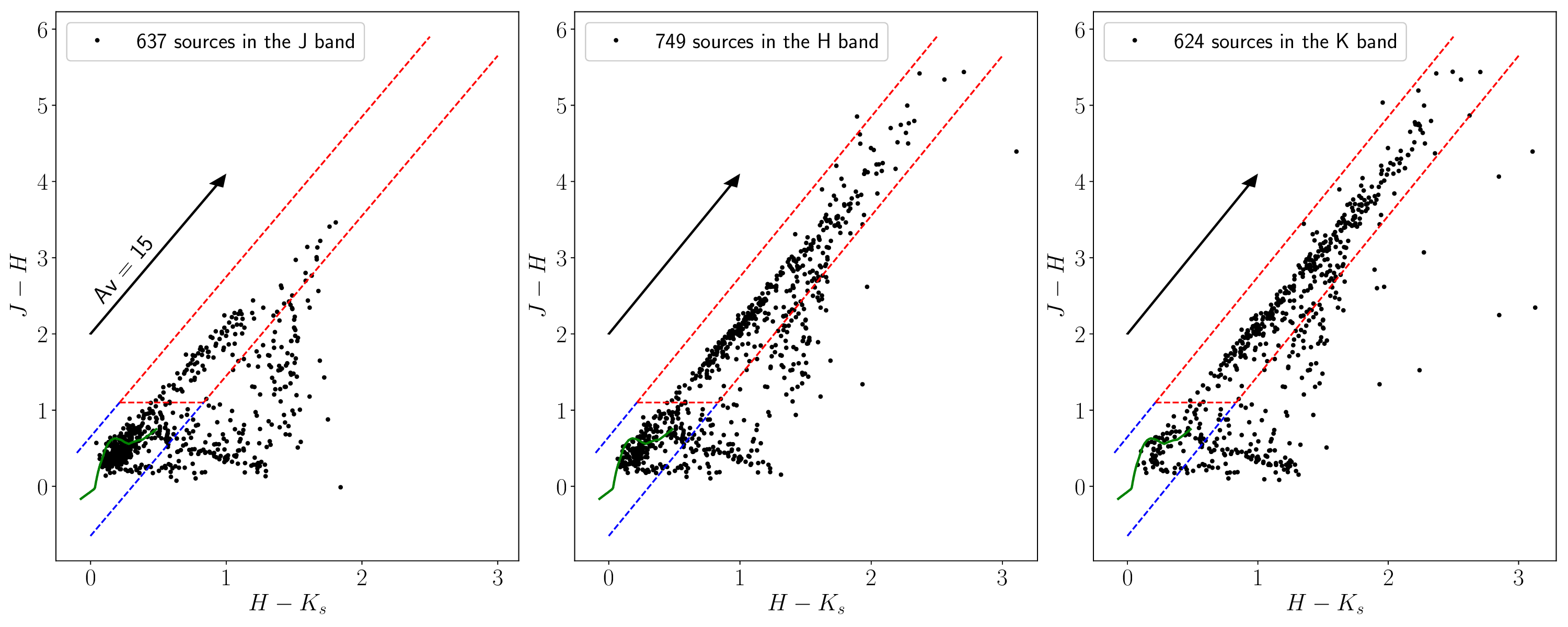}
   \caption{$J-H$ vs. $H-K_s$ colour-colour diagrams of the sources in the $JHK_s$ bands from left to right respectively. The main-sequence intrinsic colours (highlighted in green) are adopted from \citet{2013ApJS..208....9P}. The background and foreground stars are lying within the areas enclosed by the red and blue dashed lines respectively. A reddening vector of length $A_V=15$ mag is displayed as the arrow. } 
   \label{Fig:ccd_jhk}
 \end{figure*}

\begin{table*}
	\centering
	\caption{The foreground-corrected polarisation in the $K_s$ band and VVVX $JHK_s$ photometry of the 364 background stars towards G11.}{\label{tbl:Kpol}}
	\begin{threeparttable}
	\begin{tabular}{ccccccccccccc}
		
		\hline\hline
		ID & R.A. & Decl. & $P_K$ & $\delta P_K$ & $\theta_{PA}$ & $\delta \theta_{PA}$ & J & Jerr & H & Herr & K & Kerr \\
		& (J2000)   & (J2000)   & (\%)     & (\%)    & ($\degr$) & ($\degr$) & (mag)  &  (mag)& (mag)  & (mag) & (mag)  & (mag) \\
		\hline
1 & 272.51927 & -19.52159 & 2.1 & 0.1 & 27.0 & 1.0 & 14.900 & 0.004 & 12.004 & 0.001 & 10.395 & 0.001 \\
2 & 272.48228 & -19.51889 & 1.7 & 0.1 & 85.4 & 1.9 & 14.393 & 0.003 & 12.118 & 0.001 & 10.927 & 0.002 \\
3 & 272.54558 & -19.51703 & 4.3 & 0.1 & 43.8 & 0.8 & 17.315 & 0.029 & 13.329 & 0.003 & 11.383 & 0.002 \\
4 & 272.46133 & -19.51687 & 4.4 & 0.4 & 19.3 & 2.8 & 18.563 & 0.088 & 14.320 & 0.008 & 12.303 & 0.005 \\
5 & 272.50304 & -19.51593 & 1.9 & 0.8 & 60.0 & 10.3 & 16.187 & 0.011 & 13.773 & 0.005 & 12.599 & 0.006 \\
6 & 272.55243 & -19.51442 & 3.3 & 0.2 & 32.9 & 1.9 & 17.374 & 0.030 & 13.673 & 0.004 & 11.878 & 0.003 \\
7 & 272.47824 & -19.51363 & 1.0 & 0.2 & 61.2 & 4.1 & 15.065 & 0.005 & 12.595 & 0.002 & 11.448 & 0.002 \\
8 & 272.49046 & -19.51009 & 1.1 & 0.4 & 113.6 & 12.2 & 16.133 & 0.011 & 13.597 & 0.004 & 12.390 & 0.005 \\
9 & 272.57179 & -19.50984 & 3.5 & 0.1 & 18.8 & 0.5 & 15.717 & 0.008 & 12.062 & 0.001 & 10.338 & 0.001 \\
10 & 272.54242 & -19.50795 & 2.6 & 0.2 & 41.7 & 1.9 & 14.667 & 0.004 & 12.544 & 0.002 & 11.548 & 0.002 \\
\hline
	\end{tabular}
	\begin{tablenotes}
	\footnotesize
	 \item Only a portion of this table is shown here. The full version is available online \citep{G11_table}.
	\end{tablenotes}
	\end{threeparttable}
\end{table*}

\subsection{Analysis of Reddened Background Stars}
The IRSF observations are not deep enough to detect sufficient background stars, especially in the $J$ band. Similar to the approach used in \citet{2017ApJ...838...80C,2022_chen_rcw120}, we combine the $JHK_s$ starlight polarisation ($P_\lambda \geqslant 2\delta P_\lambda$) with the VVVX photometry. Applying a search radius of $1\arcsec$, there are 637 sources in the $J$ band, 749 sources in the $H$ band, and 624 sources in the $K_s$ band, that have starlight polarisation $P_\lambda \geqslant 2\delta P_\lambda$ and $JHK_s$ photometry from VVVX. The VVVX sources with starlight polarisation have a mean FWHM of $0\farcs9$, a value not too much smaller than the mean FWHM ($\sim1\farcs5-1\farcs8$) of the IRSF observations. Although the IRSF observations have larger FWHM than the mean FWHM of VVVX sources, the cross-match between the starlight polarisation and VVVX catalog is mostly one-to-one. In the $K_s$ band sample of 624 sources with polarisation and VVVX photometry, as an example, only one source has another fainter source from the VVVX catalog within a $2\arcsec$ radius. The fractions of blended sources are believed to be very low for the starlight polarisation samples in the $JHK_s$ bands.

We applied the classical $J-H$ versus $H-K_s$ colour-colour diagram to classify background stars that mostly show large $J-H$ and $H-K_s$ colours. Figure~\ref{Fig:ccd_jhk} shows the colour-colour diagram for the starlight polarisation with the VVVX photometry in the $JHK_s$ bands. Two populations exist in the colour-colour diagrams, stars with low $H-K_s$ and $J-H$ colours, and those reddened to higher colours along the extinction line. Similar to the criteria in \citet{2022_chen_rcw120}, the criteria for background stars with respect to G11 are
\begin{align}
  J-H & \leqslant 2.1 \times (H-K_s) + 0.65 \\
  J-H & \geqslant 2.1\times (H-K_s) - 0.65 \\
  J-H & \geqslant 1.1 , 
\end{align}
and are displayed as the red dashed lines in Figure~\ref{Fig:ccd_jhk}. Stars enclosed by the blue dashed lines are classified as the foreground stars with respect to G11. The $J-H\geqslant1.1$ threshold for classifying background stars is a compromise value neither too high nor too low, and is chosen mainly based on our experience and the source distributions in the colour-colour diagram in Figure~\ref{Fig:ccd_jhk}. Because the distance to G11 is 3.6 kpc, 2 times that to the RCW 120 region \citep{2022_chen_rcw120}, the interstellar extinction caused by diffuse ISM towards G11 is likely higher than that for RCW 120. The higher interstellar extinction by diffuse ISM is represented by the $J-H\geqslant 1.1$ criterion for G11, higher than $J-H\geqslant0.9$ applied for RCW 120 \citep{2022_chen_rcw120}. The numbers of foreground stars are 341, 232, 86 in the $J$, $H$ and $K_s$ bands, respectively. In contrast, the numbers of background stars are 97, 334, and 364 in the $J$, $H$ and $K_s$ bands, respectively. In the short wavelengths, starlight polarisation is dominated by foreground stars. The numbers of background stars with polarisation in more than two bands are largely reduced. For example, 60 background stars meet the criterion, $P_\lambda \geqslant 2\delta P_\lambda$, in all the three bands. The number of background stars with polarisation of $P_\lambda \geqslant 2\delta P_\lambda$  in both $H$ and $K_s$ bands is 237, a value much lower than that in the $H$ or $K_s$ band. 

The Gaia mission provides accurate stellar parallaxes up to several kilo parsecs \citep{2016A&A...595A...1G,2021A&A...649A...1G}, depending on the level of interstellar extinction along the LOS. From the star distance catalog \citep{2021AJ....161..147B}, we retrieved distance estimates of $JHK_s$ starlight polarisation by using a match radius of $1\arcsec$. We considered the sources with narrow range of distance estimates, defined by $(B_{rgeo} - b_{rgeo})/rgeo < 0.2$, where $B_{rgeo}$, $b_{rgeo}$, $rgeo$ are the 84th percentile, the 16th percentile, and the median value of the geometric distance posterior \citep{2021AJ....161..147B}. 

The fraction of foreground stars with available distance estimates from Gaia is between $1/3$ to $1/2$ of all foreground stars detected in the $JHK_s$ bands. We found no background stars observed in the $JHK_s$ bands with accurate distance estimates, indicating that the observed background stars are most likely too far and/or highly reddened \citep[e.g.][]{2013A&A...557A..51C}. A robust classification of background stars relative to G11 relies on distance. Nevertheless, the smaller extinction in the $K_s$ band would, therefore, lead to more stars detected behind G11. In contrast, fewer stars behind G11 are detectable in the $J$ and $H$ bands, which have relatively larger extinction. Background stars observed in the $K_s$ band are generally more distant and more likely to be lying behind G11. In fact, background stars observed in the $K_s$ band (364 stars) are comparatively more than those detected in the $J$ and $H$ bands (97 and 334 stars, respectively). The starlight polarisation of background stars in the $K_s$ band is, therefore, more likely to be resulting from the dust grains within G11. In this paper, we analyse the $K_s$-band starlight polarisation of 364 background stars. 

\begin{figure*} 
	\centering
	\includegraphics[width=0.99\textwidth]{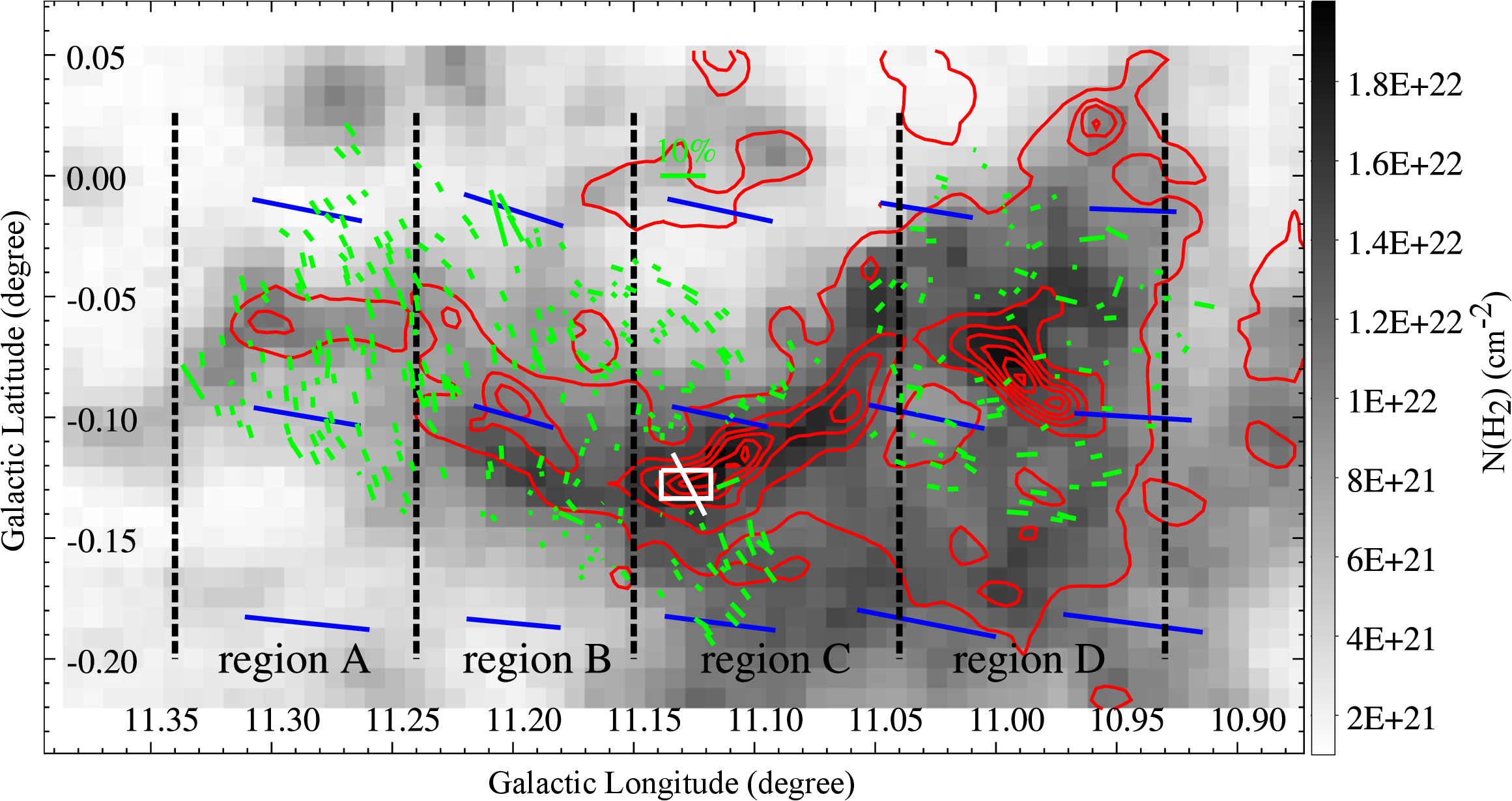}
		\caption{Molecular hydrogen column density derived from $^{13}$CO $J$=1-0 emission in the velocity range 25--37 $\mathrm{km\,s^{-1}}$, overlaid with the red contours of $N(\mathrm{H_2})_\mathrm{dust}$ from $2\times10^{22}\,\mathrm{cm^{-2}}$ to $8\times10^{22}\,\mathrm{cm^{-2}}$ in steps of $10^{22}\,\mathrm{cm^{-2}}$. The foreground-corrected $K_s$-band starlight polarisation is displayed as the green segments with lengths in proportion to polarisation degree. The blue segments with a grid spacing of $5\farcm1$ represent the magnetic field orientations, rotated $90\degr$ from the Planck 353 GHz dust polarisation angle. The mean degree of 353 GHz dust polarisation is 1.6\%. The length of the blue segment is in an arbitrary scale. The white rectangle outlines the region for which JCMT $850\,\mu$m dust polarisation was analyzed \citep{2015ApJ...799...74P}. The mean orientation of $850\,\mu$m dust polarisation is displayed as the white segment. }
		\label{Fig:NH2_KPOL}
	\end{figure*}

\subsection{Foreground polarisation}
The interstellar polarisation arising from foreground dust in front of  G11.11-0.12 is not negligible and therefore should not be ignored. For the 39 foreground stars with distances $<3\,\mathrm{kpc}$, no trend is found for the $K_s$-band polarisation degree with respect to distance. The weighted  $K_s$-band mean polarisation degree of these 39 foreground stars is $1.3\pm0.1\%$. Those contributions should be removed from the starlight polarisation of background stars prior to the analyses of magnetic fields. This correction of foreground polarisation has been widely applied when the foreground polarisation is significant and ordered \citep[e.g.][]{2013AJ....145...74C,2019ApJ...875...64E,2021ApJ...911...81D}. The foreground polarisation can be neglected in cases where it is small and random \citep[e.g.][]{2017ApJ...838...80C}.

Foreground polarisation towards G11 in the $K_s$ band was computed from the starlight polarisation of 86 foreground stars. The spatial distributions of foreground polarisation are split into grids with a spacing of $5\arcmin$. The weighted Stokes mean values of the foreground contribution in a grid is computed from the foreground polarisation within the same grid. Every background star is assigned to a nearest grid with computed foreground polarisation. The corrected Stokes values of background stars are $Q_\mathrm{cor} = Q_\mathrm{bg} - Q_\mathrm{fg}$ and $U_\mathrm{cor} = U_\mathrm{bg} - U_\mathrm{fg}$. The foreground-corrected polarisation degree ($P_K$), and position angle ($\theta_\mathrm{PA}$) are calculated using Equations (6) and (7) in \citet{2022_chen_rcw120}. The foreground-corrected starlight polarisation of the 364 background stars in the $K_s$ band are used to trace the local $B_\mathrm{sky}$ of G11. Table~\ref{tbl:Kpol} lists the foreground-corrected polarisation in the $K_s$ band and VVVX photometry in $JHK_s$-bands of the 364 background stars. The values of $\theta_\mathrm{PA}$ in Table~\ref{tbl:Kpol} are read counterclockwise from the equatorial north, following the IAU definition.

\begin{figure*} 
	\centering
	\includegraphics[width=0.99\textwidth]{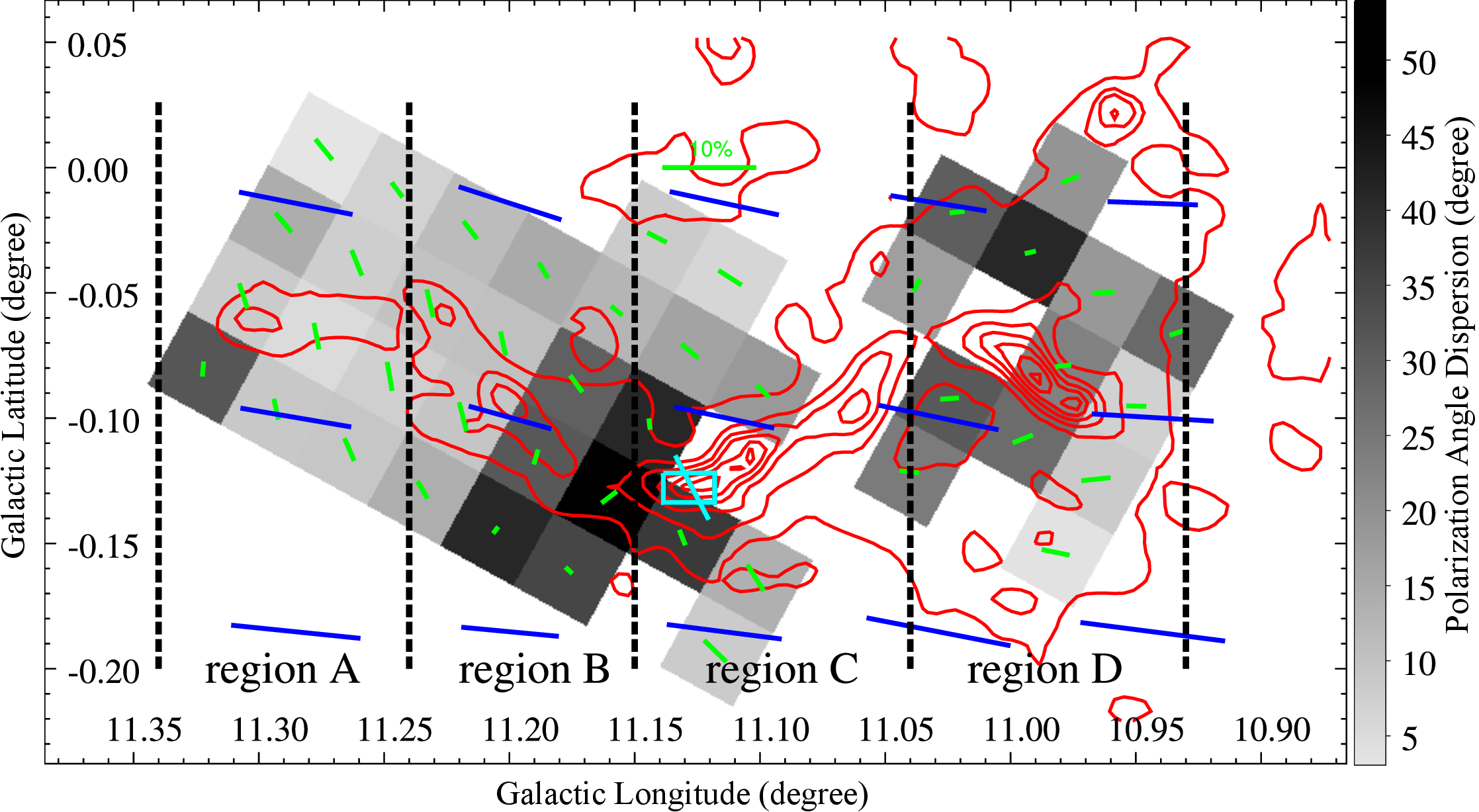}
		\caption{Polarisation angle dispersion map with a grid spacing of $2\arcmin\times2\arcmin$. The mean polarisation in each cell are displayed as the green segments with lengths in proportion to the mean polarisation degree. The blue segments, red contours, cyan rectangle and segment are the same as in Figure~\ref{Fig:NH2_KPOL}.} 
		\label{Fig:NH2_KPOL_grid}
	\end{figure*}
	
\begin{figure*} 
	\centering
	\includegraphics[width=0.99\textwidth]{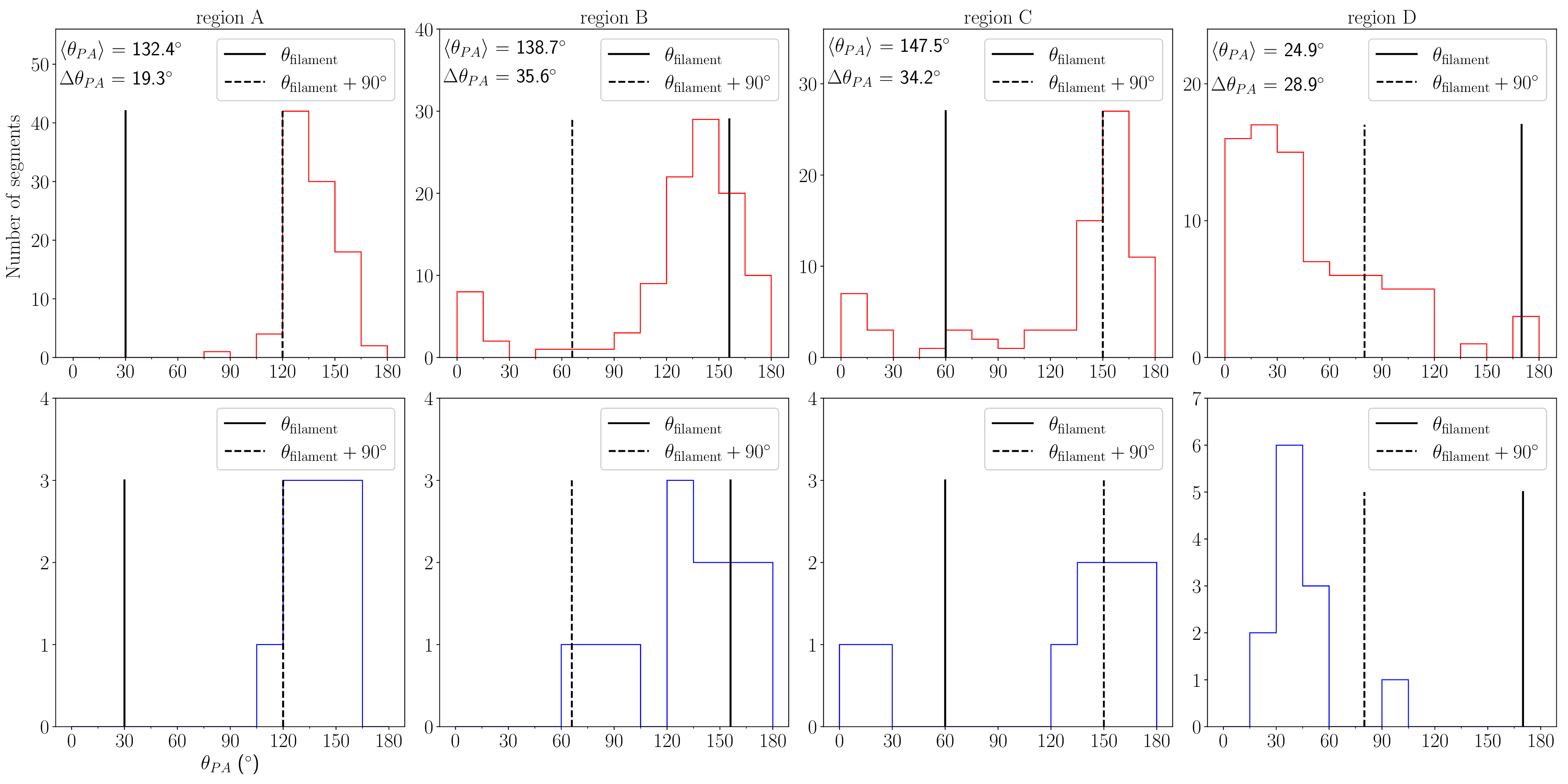}
		\caption{Top: Histograms of the orientations of the foreground-corrected $K_s$-band polarisation in the four regions of G11. The orientations of the spines in the four regions are marked by the vertical black solid lines. The orthogonal orientations of the spines are marked by the vertical dashed lines. Bottom: Same as top, but for the mean orientation in each cell. } 
		\label{Fig:PA_region}
	\end{figure*}

\subsection{The Local $B_\mathrm{sky}$ in G11}
Figure~\ref{Fig:NH2_KPOL} shows the orientation of the foreground-corrected $K_s$-band polarisation (green segments) of the background stars superimposed on the $N(\mathrm{H_2})$ map derived from $^{13}$CO emission. The $K_s$-band polarisation segments were rotated by a specific angle to match the Galactic coordinate frame of Figures~\ref{Fig:NH2_KPOL}, \ref{Fig:NH2_KPOL_grid}, \ref{Fig:Alfven_mach} and \ref{Fig:Vlsr}. Besides the polarization orientation of individual background stars shown in Figure~\ref{Fig:NH2_KPOL}, we also split the starlight polarisation into each cell of $2\arcmin\times2\arcmin$ size. Table~\ref{tbl:grid} lists the properties of each cell, including positions of the cell center, weighted mean polarisation (degrees and orientations), average gas velocity, the absolute difference between polarisation orientation and the filament orientation, and the number of polarisation vectors. We keep only the cells with at least three polarisation vectors. The $2\arcmin\times2\arcmin$ cell is moderate, so that each effective cell contains eight starlight polarisation in average. In Figure~\ref{Fig:NH2_KPOL_grid} the mean orientation of starlight polarisation in each cell was displayed as green segments, overlaid on the map of polarisation angle dispersion in each cell.

\begin{table*}
\caption{Mean starlight polarisation and mean gas velocity dispersion in each cell of $2\arcmin\times2\arcmin$ size.}{\label{tbl:grid}}
\begin{threeparttable}
\begin{tabular}{cccccccc}
\hline\hline\\
Region & R.A.& Decl. & $\langle P_K \rangle$ & $\langle \theta_{PA} \rangle$ & $\delta v$ & $|\langle\theta_{PA}\rangle -\theta_\mathrm{filament}| $ & $N_\mathrm{pol}$ \\
       & (J2000)  & (J2000) & (\%) & ($\degr$) & ($\mathrm{km\,s^{-1}}$) & ($\degr$) & \\
\hline
A & 272.6600 & -19.1720 & 2.9 & 135.5 & 1.5 & 74.5 & 6 \\
A & 272.6600 & -19.2053 & 3.0 & 128.1 & 1.7 & 81.9 & 10 \\
A & 272.6600 & -19.2387 & 3.2 & 127.1 & 1.7 & 82.9 & 11 \\
A & 272.6247 & -19.2053 & 3.0 & 139.2 & 1.4 & 70.8 & 15 \\
A & 272.6953 & -19.1720 & 1.7 & 112.6 & 1.6 & 82.6 & 11 \\
A & 272.6953 & -19.2053 & 2.3 & 129.6 & 1.8 & 80.4 & 12 \\
A & 272.6248 & -19.1720 & 2.7 & 157.8 & 2.3 & 52.2 & 9 \\
A & 272.6953 & -19.2387 & 2.7 & 140.5 & 1.9 & 69.5 & 15 \\
A & 272.5894 & -19.2053 & 2.0 & 154.3 & 2.2 & 55.7 & 4 \\
A & 272.5895 & -19.1720 & 2.9 & 156.9 & 2.2 & 53.1 & 4 \\
\hline
B & 272.6247 & -19.2387 & 3.1 & 130.3 & 1.6 & 25.7 & 7 \\
B & 272.6600 & -19.2720 & 3.3 & 131.2 & 1.8 & 24.8 & 7 \\
B & 272.6600 & -19.3053 & 1.7 & 101.8 & 1.8 & 54.2 & 8 \\
B & 272.6600 & -19.3387 & 2.1 & 67.5 & 1.7 & 88.5 & 8 \\
B & 272.6247 & -19.2720 & 2.7 & 129.3 & 1.8 & 26.7 & 11 \\
B & 272.6247 & -19.3053 & 2.2 & 153.9 & 1.8 & 2.1 & 10 \\
B & 272.6953 & -19.2720 & 2.2 & 147.2 & 1.8 & 8.8 & 4 \\
B & 272.6953 & -19.3053 & 0.9 & 81.9 & 2.0 & 74.1 & 5 \\
B & 272.6953 & -19.3387 & 1.0 & 165.0 & 2.1 & 9.0 & 10 \\
B & 272.5894 & -19.3053 & 1.5 & 168.0 & 2.2 & 12.0 & 14 \\
B & 272.5894 & -19.2720 & 2.0 & 148.8 & 1.9 & 7.2 & 11 \\
B & 272.5894 & -19.2387 & 2.4 & 155.1 & 2.0 & 0.9 & 9 \\
\hline
C & 272.6600 & -19.3720 & 1.8 & 138.4 & 1.8 & 78.4 & 5 \\
C & 272.6247 & -19.3387 & 1.2 & 124.1 & 1.9 & 64.1 & 11 \\
C & 272.6600 & -19.4053 & 3.3 & 149.3 & 1.9 & 89.3 & 13 \\
C & 272.5893 & -19.3720 & 1.8 & 164.2 & 2.1 & 75.8 & 11 \\
C & 272.5894 & -19.3387 & 2.3 & 168.3 & 2.8 & 71.7 & 14 \\
C & 272.6953 & -19.4053 & 3.3 & 163.9 & 2.0 & 76.1 & 4 \\
C & 272.5893 & -19.4387 & 2.1 & 24.0 & 1.9 & 36.0 & 7 \\
C & 272.5540 & -19.3387 & 2.9 & 176.1 & 2.5 & 63.9 & 4 \\
C & 272.5541 & -19.3053 & 2.3 & 1.9 & 2.5 & 58.1 & 3 \\
\hline
D & 272.5539 & -19.4387 & 2.0 & 34.5 & 2.0 & 44.5 & 9 \\
D & 272.5539 & -19.4720 & 2.4 & 52.5 & 2.0 & 62.5 & 7 \\
D & 272.5186 & -19.4720 & 1.7 & 37.1 & 2.1 & 47.1 & 5 \\
D & 272.5185 & -19.5053 & 2.2 & 28.9 & 2.5 & 38.9 & 4 \\
D & 272.4832 & -19.4720 & 2.4 & 33.2 & 2.5 & 43.2 & 5 \\
D & 272.4832 & -19.4387 & 1.2 & 44.1 & 2.0 & 54.1 & 10 \\
D & 272.5186 & -19.4053 & 1.6 & 91.2 & 2.0 & 78.8 & 6 \\
D & 272.5892 & -19.5053 & 3.0 & 18.3 & 2.0 & 28.3 & 5 \\
D & 272.5539 & -19.5053 & 3.2 & 36.7 & 2.3 & 46.7 & 6 \\
D & 272.4832 & -19.5053 & 1.6 & 48.9 & 2.7 & 58.9 & 6 \\
D & 272.4479 & -19.4387 & 2.0 & 52.2 & 2.1 & 62.2 & 3 \\
D & 272.4833 & -19.4053 & 1.6 & 37.0 & 2.1 & 47.0 & 6 \\
\hline
\end{tabular}
	\begin{tablenotes}
	
	\item The absolute difference between mean orientation of polarisation in each cell and the filament orientation is presented in the column of $|\langle\theta_{PA}\rangle -\theta_\mathrm{filament}| $. The number of starlight polarisation in each cell is listed in the column of $N_\mathrm{Pol}$.
	\end{tablenotes}
\end{threeparttable}
\end{table*}

The spine of G11, better traced by the $N(H_2)_{dust}$ contours, clearly shows a positive gradient of column density, from left to right. The spine of G11 is, in fact, twisted along the Galactic plane, and can be divided into four regions with different orientations relative to the Galactic plane. The spine orientations of the four regions, read counterclockwise from the equatorial north, were determined by the visual inspection of the highest $N(H_2)_{dust}$ isocontours in each region.

Region A:  $11\fdg24<l<11\fdg34$; the spine here has the lowest $N(H_2)_{dust}$ of $\sim2\times10^{22}\,\mathrm{cm^{-2}}$, with a peak value of $3.6\times10^{22}\,\mathrm{cm^{-2}}$. The envelope of region A has even lower column density, $\sim 0.5\times10^{22}\,\mathrm{cm^{-2}}$, derived from the $^{13}$CO $J$=1-0 emission. 

Region B: $11\fdg15<l<11\fdg24$; the spine of region B has a peak $N(H_2)_{dust}$ of $4.7\times10^{22}\,\mathrm{cm^{-2}}$, higher than that of region A. The envelope here has column density $\sim0.8\times\gtrsim 10^{22}\,\mathrm{cm^{-2}}$. 

Region C: $11\fdg04<l<11\fdg15$; the spine here contains the densest and most massive dust condensation, P1 \citep{2003ApJ...588L..37J}, in which massive star formation is evident \citep[e.g.][]{2006A&A...447..929P,2014A&A...570A..51S,2014ApJ...796..130R}. The peak $N(H_2)_{dust}$ reaches $7.7\times10^{22}\,\mathrm{cm^{-2}}$ towards the dust condensation P1. The magnetic fields in the spine of this region are perpendicular to the filament itself \citep{2015ApJ...799...74P}, as indicated by the white segment in Figure~\ref{Fig:NH2_KPOL}. 

Region D: $10\fdg93<l<11\fdg04$; the spine of region D has a peak $N(H_2)_{dust}$ of $8.3\times10^{22}\,\mathrm{cm^{-2}}$ towards its densest part, and is almost as dense as that of region C. It contains the massive dust condensation P6 \citep{2003ApJ...588L..37J,2014MNRAS.439.3275W} and the $70\,\mu$m dark clump  G10.99--0.08 \citep{2019A&A...622A..54P,2019ApJ...886..102S}, in which star formation is still on-going.

The orientations of the foreground-corrected $K_s$-band polarisation in the four regions of G11 are displayed in the top rows of Figure~\ref{Fig:PA_region}. The orientations of mean starlight polarisation in cells in the four regions are shown in the bottom rows.
In regions A, B and C, the overall orientations of starlight polarisation are in the range $130-150\degr$. The starlight polarisation in region A shows a narrow distribution, with a peak at $132\degr$. The orientations of starlight polarisation in region B are similar to those in region A, but show a larger scatter. In areas of region B adjacent to region A, the orientations of starlight polarisation are parallel to those in region A. The orientations of starlight polarisation in region C have a peak at $147\degr$, roughly parallel to those in regions A and B. The orientations of starlight polarisation in region D peak at $32\degr$, roughly perpendicular to those in regions A, B and C. 

The cutout 353 GHz Stokes $I, Q, U$ maps with pixel scale of $1\farcm7$ for G11 were retrieved from the Planck Legacy Archive \footnote{http://http://pla.esac.esa.int/pla}. The Planck $I, Q, U$ maps were smoothed to $5\farcm1$ resolution using a Gaussian kernel with FWHM = 3 pixels. The 353 GHz dust polarisation degree and polarisation orientation in the IAU definition were computed from the Stokes $I, Q, U$ maps, following Equations 1 and 2 in \citet{2015A&A...576A.104P}. The magnetic field orientations on the plane of the sky are displayed as the blue segments in Figures~\ref{Fig:NH2_KPOL} and \ref{Fig:NH2_KPOL_grid}.
The magnetic field orientations, rotated $90\degr$ from the Planck dust polarisation angle, are compared with the field orientations traced by the foreground-corrected starlight polarisation angles. The average offsets in magnetic field orientations traced by the two methods are $68\degr$, $57\degr$, $50\degr$, and $4\degr$ in regions A, B, C, and D respectively, an indication that magnetic fields traced by starlight polarisation are very different from the field orientations traced by dust polarisation in most parts of the G11 filament. G11 is lying within the Galactic plane, along which direction the Planck dust polarisation data record all the polarised dust emission along the same LOS. The LOS contamination might be much less severe for the filaments at high Galactic latitude. \citet{2016A&A...596A..93S} compared the orientation of magnetic field projected on the plane of the sky inferred from the Planck 353 GHz dust polarisation and optical/infrared starlight polarisation for four nearby molecular clouds at high Galactic latitude. The consistency between field orientation sampled in the dust polarised emission and dust extinction suggests considerable agreement between the magnetic field orientation on the plane of the sky estimated with both techniques at comparable scales \citep{2016A&A...596A..93S}. Because of the severe LOS contamination, the Planck dust polarised emission towards G11 are inadequate in tracing the local magnetic field of this filament.

Figure~\ref{Fig:PA_region} compares the polarisation orientations with the spine orientations in the four regions. In both regions A and C, the orientations of starlight polarisation are perpendicular to their spines.
The mean starlight polarisation in the cells within region B show a bimodal distributions. In areas of region B adjacent to region A, the starlight polarisation is roughly parallel to the spine of region B. In the rest parts of region B, the grid-averaged starlight polarisation tends to be perpendicular to the spine of region B, but with higher dispersion, as shown by Figure~\ref{Fig:NH2_KPOL_grid}. The projection effects of the magnetic field and filaments can produce small relative orientation angles between the filament and the magnetic field projected onto the plane of the sky, even though the magnetic field is orthogonal to the filament in a 3D space \citep{2020ApJ...899...28D}. The observed parallel orientation between the starlight polarisation and the spine in region B is likely to be due to the projection effects. The orientations of starlight polarisation in region D are almost perpendicular to those in regions A, B and C. However, the grid-averaged starlight polarisation in region D is close to being perpendicular to the spine of region D. The orientations of the foreground-corrected $K_s$-band polarisation in the four regions are generally perpendicular to the spines in each region, despite the fact that the spines in each region have different orientations. 

The $N(\mathrm{H_2})_\mathrm{dust}$ in the spine of region C is $\gtrsim4\times10^{22}\,\mathrm{cm^{-2}}$. Few starlight polarisation was detected towards the spine of region C. Starlight polarisation traces better the $B_\mathrm{sky}$ in the surrounding envelope of region C. The $B_\mathrm{sky}$ traced by starlight polarisation in the envelope is consistent with the magnetic fields on the spine of region C obtained from dust polarisation \citep{2015ApJ...799...74P}. The consistency of $B_\mathrm{sky}$ in the envelope of parsec scales, and the dense core of sub-parsec scales, indicates that magnetic fields are perpendicular to the spine of region C. Combining the near-IR starlight polarisation in this work and the dust polarisation in \citet{2015ApJ...799...74P}, the magnetic fields are found to be always perpendicular to G11 across the parsec-scale envelope surrounding the filament to the dense spine in which star formation is on-going.

\begin{table*}
\centering
\caption{The physical properties of the four regions in G11. }\label{tbl:parts}
\begin{threeparttable}
\begin{tabular}{ccccccc}
\hline\hline
   Region  & $n(\mathrm{H_2})$ & $\delta v$ & $\delta \phi$ & $B_\mathrm{sky}$ & $v_\mathrm{A}$ & $\mathcal{M}_\mathrm{A}$ \\
         & ($10^{3}\,\mathrm{cm^{-3}}$) & ($\mathrm{km\,s^{-1}}$) & ($\degr$)  & ($\mu$G) & ($\mathrm{km\,s^{-1}}$)  & \\
   \hline
   A & 0.4  & 1.8   & $7\pm1$  & 108 (54) & 6.3 (3.2) & 0.3 (0.6) \\ 
   B & 0.5  & 1.9   & $20\pm4$  & 48 (40) & 2.8 (2.3) & 0.7 (0.8) \\
    C & 0.7  &  2.0  & $22\pm3$  & 55 (47) &  3.2 (2.7) & 0.6 (0.7) \\
   D  & 0.8  & 2.1 & $33\pm9$    & 39 (42) &  2.3 (2.4) & 0.9 (0.9) \\
  \hline
\end{tabular}
	\begin{tablenotes}
	
	\item The field strength, Alfv{\'e}n speed, and Alfv{\'e}n Mach number, estimated from the DCF and ST methods are both given in the columns of $B_\mathrm{sky}$, $v_\mathrm{A}$, and $\mathcal{M}_\mathrm{A}$. The ST estimates are always in the parentheses after the DCF estimates. 
	\end{tablenotes}
\end{threeparttable}
\end{table*}

\subsection{Strength of $B_\mathrm{sky}$}
The variable extinction across G11 leads to the non-uniform spatial distribution of starlight polarisation. The number of starlight polarisation towards the low-density envelope of G11 is much higher than that towards the high-density spine. Therefore, the magnetic field strength is more representative of the envelope rather than the dense spine of G11. \citet{2015ApJ...799...74P} estimated a total magnetic strength of $>267\pm26\,\mu$G for the dense condensation in region C (the rectangle in Figures~\ref{Fig:NH2_KPOL} and \ref{Fig:NH2_KPOL_grid}). In the envelope of G11, the gas density is much lower than that in the spine. Assuming energy equipartition between turbulence and perturbed magnetic fields, given by 
\begin{align}
\frac{1}{2}\rho \delta v^2\sim \frac{ \delta B^2}{8\pi},    
\end{align}
the polarisation angle dispersion, $\delta \phi\approx \frac{\delta B}{B}$, yields the Davis-Chandrasekhar-Fermi (DCF) method
\citep{1951ApJ...114..206D,1953ApJ...118..113C} for estimating the on-sky strength of magnetic field
\begin{align}\label{equ:dcf}
    B_\mathrm{sky} = Q \sqrt{4\pi \rho }\frac{\delta \nu}{\delta \phi},
\end{align}
where $Q$ is the correction factor, $\rho$ is the gas density, and $\delta v$ is the gas velocity dispersion.  

The $^{13}$CO $J$=1-0 emission is mostly optically thin within the envelope of G11. The moment 2 map of $^{13}$CO $J$=1-0 emission provides the velocity dispersion $\delta v$ of molecular gas across G11. The median $\delta v$ values of the areas overlapping with starlight polarisation are extracted from the moment 2 map of $^{13}$CO emission. The density of molecular gas, or molecular hydrogen number density $n(\mathrm{H_2})$ multiplied by $\mu m_\mathrm{H}$ ($\mu=2.83$ is the average molecular gas weight \citep{2008A&A...487..993K}, $m_\mathrm{H}$ is the mass of a single hydrogen atom), is always a challenging task in characterizing the fundamental properties of molecular clouds. The mean gas density of G11 was estimated to be as $n(\mathrm{H_2})\approx0.5\times10^3\,\mathrm{cm^{-3}}$ at a scale of $\sim 5$\,pc  \citep[see Fig.4 in][]{2013A&A...557A.120K}. Assuming the LOS extent $\sim5$ pc for the envelope of G11, $n(\mathrm{H_2})$ is approximately $N(\mathrm{H_2}$)/5 pc, where $N(\mathrm{H_2})$ is the hydrogen column density derived from $^{13}$CO $J$=1-0 emission. The median $n(\mathrm{H_2})$ of region A is $0.4\times10^3\,\mathrm{cm^{-3}}$, roughly consistent with the large scale $n(\mathrm{H_2})\approx0.5\times10^3\,\mathrm{cm^{-3}}$ in \citet{2013A&A...557A.120K}.

The starlight polarisation orientations show ordered mean orientation in the four regions of the G11 filament, an indication of ordered magnetic fields. Therefore, the polarisation angle dispersion $\delta \phi$ due to turbulence can be determined from the angular dispersion function \citep[ADF;][]{2008ApJ...679..537F,2009ApJ...696..567H,2009ApJ...706.1504H}. The ADF method obtains a measure of the difference in polarisation angle of $N(l)$ pairs of starlight polarisation separated by displacement $l$ \citep{2009ApJ...696..567H}:
\begin{align}
\centering
\langle \Delta\Phi^2(l)\rangle^{1/2} = \bigg \{\frac{1}{N(l)}\displaystyle\sum_{i=1}^{N(l)}[\Phi(x) - \Phi(x+l)]^2 \bigg \}^{1/2}.
\end{align}
The polarisation angle dispersion attributed to the turbulent magnetic field, $B_t(x)$, and the large scale structure magnetic field, $B_0(x)$, are folded into the $\langle \Delta\Phi^2(l)\rangle^{1/2}$. If the large scale and turbulent magnetic fields are statistically independent, and the turbulent dispersion of polarisation angle is a constant $b$ for the displacement $l$ larger than the turbulent correlation length $\delta$, characterizing $B_t(x)$, the following equation holds
\begin{align}\label{equ:adf}
\centering
\langle \Delta\Phi^2(l)\rangle_\mathrm{tot} \approx b^2 + m^2l^2 + \sigma_M^2(l),
\end{align}
when $\delta  < l\ll d$, where $d$ is the typical scale length of variations in $B_0(x)$, and $\sigma_M(l)$ is the measurement uncertainty \citep{2009ApJ...696..567H}. \citet{2009ApJ...706.1504H} obtained a turbulent correlation length $\delta\approx0.016$ pc for the massive star-forming cloud OMC--1 at a distance of 450 pc. Following the same method as \citet{2009ApJ...706.1504H}, \citet{2010ApJ...723..146F} estimated the turbulent correlation length $\delta$ to be of a few 0.001 pc for the Pipe nebula at a distance of 145 pc. For the G11 filament, at a distance of 3.6 kpc, the closest separation between any pair of starlight polarisation is $\approx0.1$ pc, at least one order of magnitude higher than the turbulent correlation length $\delta$ found for OMC--1 cloud and Pipe nebula. The length of G11 is about 30 pc and the sizes of regions A, B, C and D are $\gtrsim6$ pc. The starlight polarisation towards G11 satisfies the condition $\delta  \ll l\ll d$, which implies that Equation~\ref{equ:adf} is suitable for deriving the turbulent dispersion of polarisation angle. 

\begin{figure*} 
   \centering
   \includegraphics[width=0.99\textwidth]{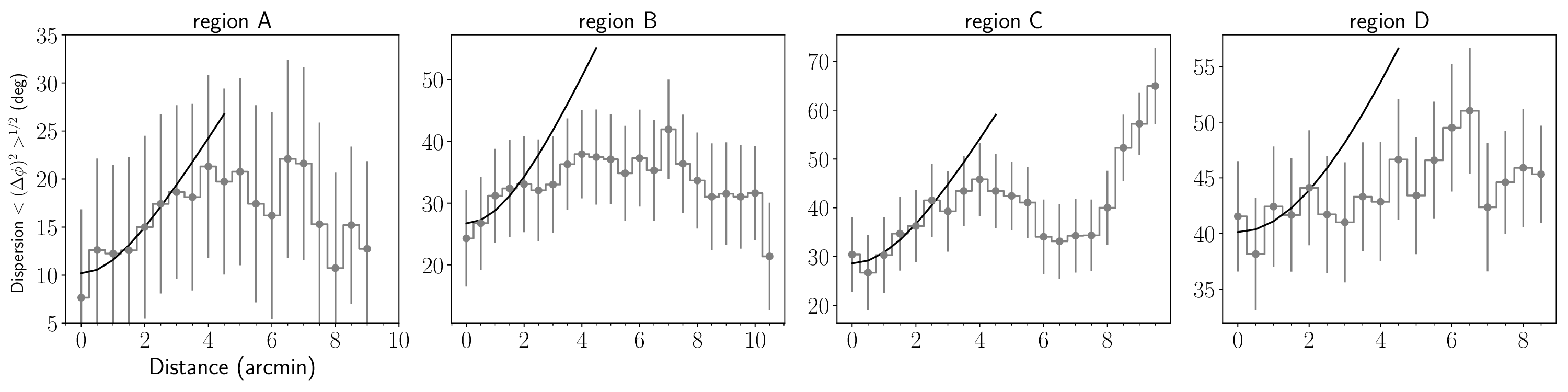}
   \caption{The angular dispersion function vs. displacement distance of the four regions in G11. The solid line represents the best fit to the ADF of the four regions. The measurement uncertainties were subtracted prior to computing the fits to the data.   } 
   \label{Fig:ADF}
\end{figure*}

\begin{figure*} 
   \centering
   \includegraphics[width=0.95\textwidth]{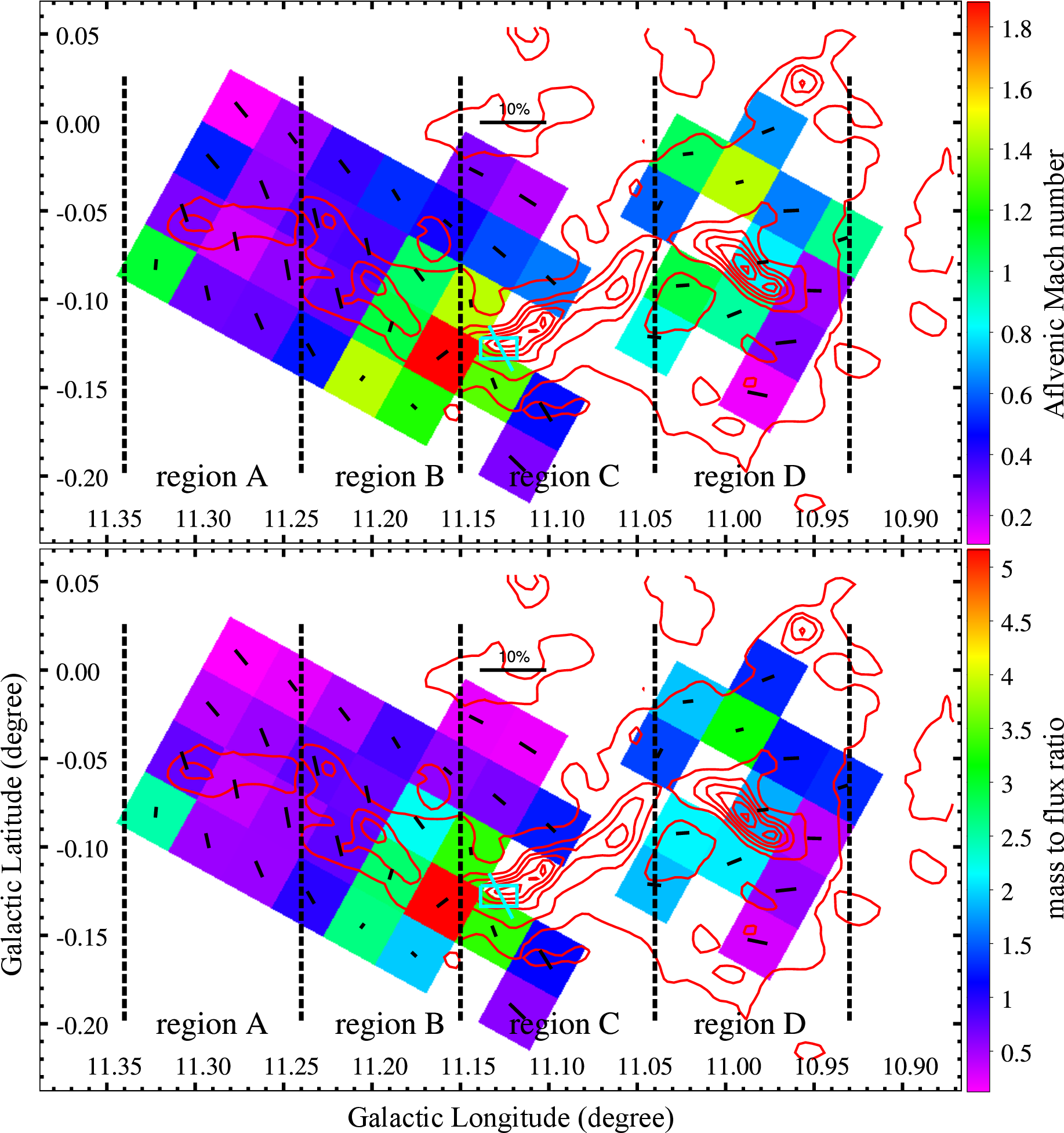}
   \caption{Alfv{\'e}n Mach number map (top panel) and dimensionless $\lambda$ map (bottom panel) of G11. The mean polarisation in each cell is as same as in Figure~\ref{Fig:NH2_KPOL_grid}, but is displayed as the black segments. The red contours, cyan rectangles and segments are as same as in Figure~\ref{Fig:NH2_KPOL_grid}.} 
   \label{Fig:Alfven_mach}
 \end{figure*}

\begin{figure*} 
   \centering
   \includegraphics[width=0.95\textwidth]{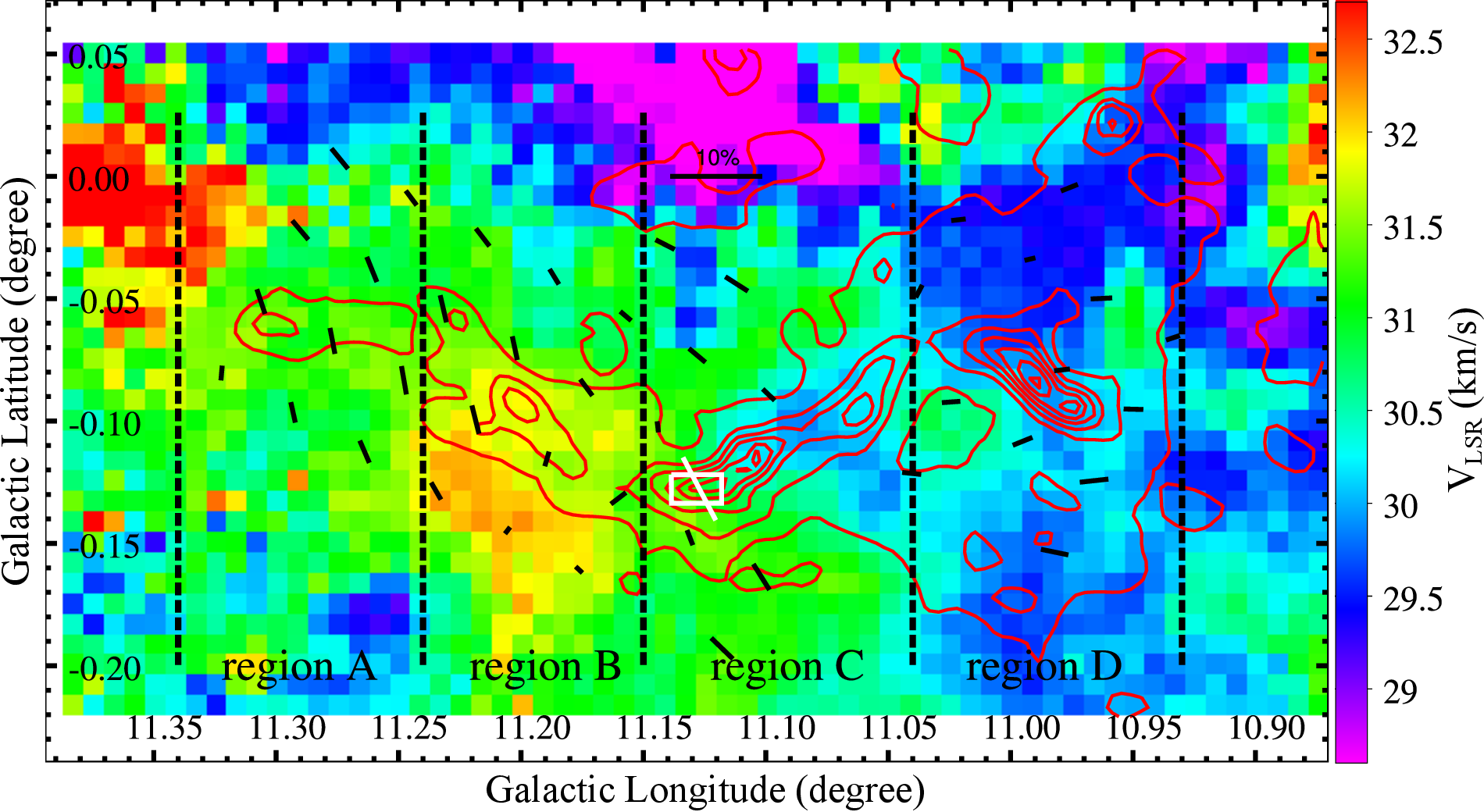}
   \caption{Moment 1 map from the $^{13}$CO $J$=1-0 emission in the velocity range of $25-37\,\mathrm{km\,s^{-1}}$. The mean polarisation in each cell is as same as in Figure~\ref{Fig:NH2_KPOL_grid}. The red contours, white rectangle and segment are as same as in Figure~\ref{Fig:NH2_KPOL}. }
   \label{Fig:Vlsr}
 \end{figure*}

Figure~\ref{Fig:ADF} shows the ADF of the four regions along with the best fit from Equation~\ref{equ:adf} using the first five data points ($l\leqslant2\farcm5$ or $\leqslant2.6$ pc). The constant $b$ is determined by the zero intercept of the best fit to the ADF, at $l=0$. According to the work by \citet{2009ApJ...706.1504H}, the ratio of turbulent to large scale magnetic fields is a function of the constant $b$:
\begin{align}\label{equ8}
\frac{\langle B_t^2\rangle^{1/2}}{B_0} = \frac{b}{\sqrt{2-b^2}}.
\end{align}

We use $\delta B$ to replace $\langle B_t^2\rangle^{1/2}$, and define the polarisation angle dispersion as $\delta \phi=\frac{\delta B}{B_0}$. Table~\ref{tbl:parts} shows the results of ADF analyses of the four regions in G11.

We utilize the $\delta \phi$ obtained from the ADF analyses to estimate the strength of $B_\mathrm{sky}$ in the four regions. MHD simulations found that the original DCF method would overestimate the field strength. Therefore, a correction factor, $Q$ in Equation~\ref{equ:dcf}, was proposed to bring the field strength derived from the DCF method closer to the model field strength \citep{2001ApJ...546..980O,2001ApJ...561..800H}. For the ordered magnetic fields with $\delta \phi < 25\degr$, $Q\approx0.5$ yields a reasonable estimate on the field strength \citep{2001ApJ...546..980O,2001ApJ...561..800H}. For regions C and D with $\delta \phi \gtrsim 25\degr$, we still assume $Q\approx0.5$, but the field strength within them is quite uncertain. From Equation~\ref{equ:dcf}, we estimate the $B_\mathrm{sky}$ strength of the four regions in G11. We present the $B_\mathrm{sky}$ strength estimated from the DCF method, and the corresponding Alfv{\'e}n speed $v_\mathrm{A}$ and Alfv{\'e}n Mach number $\mathcal{M}_\mathrm{A}$ in Table~\ref{tbl:parts} for the four regions.

The DCF method is based on the assumption that turbulence is sub-Alfv{\'e}nic. MHD simulations demonstrated that the field strength obtained by the DCF method is reliable when molecular gas is sub-Alfv{\'e}nic or equivalently $\delta \phi<25\degr$ \citep{2001ApJ...546..980O,2008ApJ...679..537F,2021ApJ...919...79L}. The field lines close to the LOS would cause large polarisation angle dispersion even if the turbulence is sub-Alfv{\'e}nic \citep{2008ApJ...679..537F}. The polarisation angle dispersion, $\delta \phi$, is smallest in region A. The highly ordered starlight polarisation in region A implies that magnetic fields in region A are close to the plane of the sky. The projection effect due to the field orientation is small in region A. The field strength in region A, estimated from the DCF method, is $108\,\mu$G, while much lower field strength is found in regions B, C, and D.

\citet{2021A&A...647A.186S} proposed a new method for estimating magnetic field strength in the compressible ISM. Unlike the DCF method assuming the cross term $\delta B\cdot B_0 = 0 $, \citet{2021A&A...647A.186S} included this cross term, and assumed the turbulent kinetic energy equal to the magnetic energy fluctuations. They obtained
\begin{align}
\frac{1}{2}\rho \delta v^2= \frac{ \delta B B_0}{4\pi},    
\end{align} and the new method (hereafter ST) for estimating magnetic field strength 
\begin{align}\label{equ:st}
 B_\mathrm{sky} =  \sqrt{2\pi \rho }\frac{\delta \nu}{\sqrt{\delta \phi}}.   
\end{align}

Although the polarisation angle dispersion $\delta \phi$ is caused by the perpendicular fluctuations of magnetic energy rather than the parallel  fluctuations, \citet{2021A&A...656A.118S} demonstrated that perpendicular fluctuations are comparable to the parallel fluctuations, thus $\delta \phi$ can also be utilized in the ST method. The magnetic field strength in the four regions, estimated by the ST method, are given in the parentheses after the DCF estimates in Table~\ref{tbl:parts}. The field strength estimated by the ST method is lower than that by the DCF method. The reduction factor is nearly two for region A, in which $\delta \phi$ is small. In regions B, C, and D with much higher $\delta \phi$ and higher density, the ST method returns field strength comparable to the DCF estimates. In contrast to the DCF method, the square root of $\delta \phi$ used in the ST method will weaken the effect of $\delta \phi$ variation on calculating the magnetic field strength. The ST method derives rather constant field strength, about $50\,\mu$G, for the four regions.

With the field strength estimated from the DCF or ST method, $v_\mathrm{A}$ in the four regions are computed using $v_\mathrm{A} = |B_\mathrm{sky}|/\sqrt{4\pi\rho}$. Table~\ref{tbl:parts} lists the $v_\mathrm{A}$ from the DCF and ST methods. The Alfv{\'e}nic Mach number $\mathcal{M}$ is a function of the polarisation angle dispersion $\delta \phi$, depending on the DCF or ST method. Following the DCF method, $\mathcal{M}_\mathrm{A}=\frac{\delta \phi}{Q}$, depends only on $\delta \phi$ when the correction factor $Q\approx0.5$ is fixed. Note that the ST method derives $\mathcal{M}_\mathrm{A}=\sqrt{2 \delta \phi}$, higher than that derived from the DCF method. For instance, $\mathcal{M}_\mathrm{A}$ of region A estimated from the ST method is 0.6, twice that estimated from the DCF method. The difference in $\mathcal{M}_\mathrm{A}$ between the two methods is small for regions B, C, and D. The turbulence within the four regions of G11 is generally sub-Alfv{\'e}nic to trans-Alfv{\'e}nic. If the turbulence is super-Alfv{\'e}nic, the isotropic turbulence would twist the field lines. The average field orientations traced by starlight polarisation would be close to random distribution, and the average starlight polarisation degree is likely to be small. The average starlight polarisation per cell (see Table~\ref{tbl:grid}) is aligned perpendicular to the G11 filament, and the average polarisation degree per cell is typically 2-3\%. The turbulence within G11 is sub-Alfv{\'e}nic to trans-Alfv{\'e}nic, indicating that magnetic fields are generally more important than, or as important as, turbulence.

\section{Discussion}
\subsection{Sub-Alfv{\'e}nic to trans-Alfv{\'e}nic envelope}
The starlight polarisation within the $2\arcmin\times2\arcmin$ grid spacing across G11, in principle, provides the information about the $\mathcal{M}_\mathrm{A}$ distribution across G11. Figure~\ref{Fig:NH2_KPOL_grid} shows the map of polarisation angle dispersion $\delta \phi$ with a grid spacing of $2\arcmin\times2\arcmin$. The $\mathcal{M}_\mathrm{A}$ map, derived from $\mathcal{M}_\mathrm{A}=\frac{\delta \phi}{0.5}$ following the DCF method, is shown in the top panel of Figure~\ref{Fig:Alfven_mach}. 
In region A, the $\mathcal{M}_\mathrm{A}$ values obtained from the ST method are roughly twice the values from the DCF method; in regions B, C, and D, the $\mathcal{M}_\mathrm{A}$ values obtained from both methods are overall consistent (see also Table~\ref{tbl:parts}). The $\mathcal{M}_\mathrm{A}$ distributions across G11 is sub-Alfv{\'e}nic to trans-Alfv{\'e}nic, an indication that magnetic field channels the gas flow from the envelope to the spine of G11. To better explore the gas flow from the envelope to the spine, we present the moment 1 map from the $^{13}$CO $J$=1-0 emission along with magnetic fields in Figure~\ref{Fig:Vlsr}. The moment 0 and moment 2 maps of $^{13}$CO $J$=1-0 emission are shown in Figures~\ref{Fig:mom0} and ~\ref{Fig:mom2} in Appendix \ref{Sect:appendix}. In region A the gas velocity is fairly uniform, indicating little velocity gradient along the LOS. The magnetic fields in region A are likely oriented on the plane of the sky, implied by the highly aligned starlight polarisation in region A. Such oriented magnetic fields channel gas flow from the envelope to the spine without producing observable velocity gradient along LOS, which is consistent with the uniform velocity field in region A. Different from region A, the LOS components of magnetic fields in region B might be more pronounced. The gas flow along the LOS components of magnetic fields from the envelope to the spine in region B would lead to observable velocity gradient perpendicular to the spine. The gas flow from the envelope in both side onto the spine in middle would cause red-shifted and blue-shifted components relative to the gas velocity in the spine. In Figure~\ref{Fig:Vlsr} we note the $32\,\mathrm{km\,s^{-1}}$ area roughly parallel to spine, the area with velocity below $31\,\mathrm{km\,s^{-1}}$ on the opposite site, and the spine with velocity in between $31-32\,\mathrm{km\,s^{-1}}$. The blue-shifted and red-shifted areas on the two sides of the spine in region B suggest that gas flow from the envelope onto the spine is threaded by the magnetic fields.

\subsection{Mass-to-flux ratio distribution}
The critical mass-to-flux ratio for the magnetic stability, $(M/\Phi)_\mathrm{crit} = \frac{1}{2\pi \sqrt{G}}$, is given by \citet{1978PASJ...30..671N}, where $G$ is the gravitational constant. The mass-to-flux ratio normalized by $(M/\Phi)_\mathrm{crit}$ is evaluated as follows
\begin{align}\label{equ9}
\lambda = \frac{M/\Phi}{(M/\Phi)_\mathrm{crit}}=2\pi \sqrt{G} \mu m_\mathrm{H} \frac{N(\mathrm{H_2})}{B_\mathrm{sky}},
\end{align}
where $\mu=2.83$ is the average molecular gas weight \citep{2008A&A...487..993K}. The $N(\mathrm{H_2})$ map used in deriving the $\lambda$ map is from the $^{13}$CO $J$=1-0 emission across G11. The field strength $|B_\mathrm{sky}|$ estimated from the DCF method and $N(\mathrm{H_2})$ maps yield the distribution of $\lambda$ across G11, as shown in the bottom panel of Figure~\ref{Fig:Alfven_mach}. We also obtained the $\lambda$ distribution across G11 based on the field strength from the ST method. The two methods show consistency for $\lambda$ in regions B, C, and D, in which field strengths estimated from both methods are comparable to each other (see also Table~\ref{tbl:parts}). The $\lambda$ distributions in regions B, C, and D are mostly $\lambda \gtrsim 1$, indicative of magnetically critical envelope in these regions of G11. When the field strength of $\sim300\,\mu$G and an average $N(\mathrm{H_2})_\mathrm{dust}\sim6.5\times10^{22}\,\mathrm{cm^{-2}}$ are adopted for the spine of region C \citep{2015ApJ...799...74P}, the $\lambda$ in the spine of region C is reduced to a level of marginally supercritical, which is consistent with the $\lambda$ obtained for region C in this work. The $\lambda$ in region A is around 0.5 from the DCF method or close to 1 from the ST method. Moreover, the $N(\mathrm{H_2})$ in region A has a mean value $\sim6\times10^{21}\,\mathrm{cm^{-2}}$. Both the $N(\mathrm{H_2})$ and $\lambda$  in region A show some consistency with the MHD simulations by \citet{2020MNRAS.497.4196S}, in which the transition from parallel to perpendicular occurs at a transition column density $\approx10^{21-21.5}\,\mathrm{cm^{-2}}$ and at a transition $\lambda$ close to 1. 

\subsection{The role of magnetic fields in G11}
The magnetic field orientations in the envelope of G11 are almost identical to that in the spine of region C \citep{2015ApJ...799...74P}. This consistency between parsec scale and sub-parsec scale magnetic field orientations suggests the important role of magnetic fields in assembling the mass of G11. The field strength in the envelope of G11 is likely in the range $\sim50-100\,\mu$G. The median $n(\mathrm{H_2})$ of region A is $0.4\times10^3\,\mathrm{cm^{-3}}$. \citet{2015ApJ...799...74P} estimated magnetic field strength of $>267\pm26\,\mu$G in the spine of region C. They adopted $n(\mathrm{H_2})\approx3\times10^4\,\mathrm{cm^{-3}}$ in the estimate of magnetic field strength. The density contrast between the spine and the envelope of G11 is $\sim75$. If adopting the relation $|B|\propto \rho ^{0.65}$ used in \citet{2012ARA&A..50...29C}, the magnetic field strength contrast between the spine and the envelope is as high as $\sim17$, much higher than the observed contrast of $\sim3-5$. The observed magnetic field contrast between the spine and the envelope of G11 does not follow the nominal scaling relationship between magnetic field strength and gas density. Magnetic fields are not significantly enhanced in the spine of G11 as material is assembled onto the spine. This can be explained by gas flow guided by the field lines. Gas flow along the field lines cannot efficiently compress magnetic fields, leading to a minor enhancement of field strength when gas flows are merging onto the filament and a shallower relation $|B|\propto\rho^{0.4}$ \citep{2015Natur.520..518L}. For G11, we estimate both the magnetic field strength and density for the envelope, called $B_\mathrm{env}$ and $\rho_\mathrm{env}$. Meanwhile, \citet{2015ApJ...799...74P} estimate both the magnetic field strength and density for the densest part of the G11 filament, called $B_\mathrm{fil}$ and $\rho_\mathrm{fil}$. The power index of the scaling relation can be derived by $\frac{\log B_\mathrm{fil} - \log B_\mathrm{env}}{\log \rho_\mathrm{fil} - \log \rho_\mathrm{env}}$. Adopting a field strength  $B_\mathrm{env}=75\,\mu$G between the DCF and ST estimates and $\rho_\mathrm{env} = 0.4\times10^3\,\mathrm{cm^{-3}}$ estimated in this work, as well as $B_\mathrm{fil}=267\,\mu$G and $\rho_\mathrm{env}=3\times10^4\,\mathrm{cm^{-3}}$ \citep{2015ApJ...799...74P}, the field strength is proportional to $\rho^{0.3}$, similar to that measured for NGC\,6334, in which the elongated gas structures at all scales are nearly perpendicular to the magnetic fields \citep{2015Natur.520..518L}. The flat scaling relation $|B|\propto n^{0.3}$ found for G11 suggests the important role of the magnetic fields in the formation of G11. Firstly, the dense spine can easily be formed without much magnetic field resistance. Secondly, the spine becomes magnetically supercritical, because the field strength is increasing slower than gas density. The spine could reach a high volume density, resulting in a small thermal Jeans mass that is conducive to forming low-mass stars \citep{2010ApJ...720L..26L}. The outflows, traced by high-velocity CO $J$=2-1 emission towards the dust condensation G10.99--0.08 in region D \citep{2019ApJ...886..102S}, are mostly driven by low-mass protostars with masses below their mass sensitivity of $\sim1-2\,M_\odot$ \citep{2019A&A...622A..54P}. This overabundance of low-mass protostars in the densest region D of G11, and the perpendicular magnetic fields as shown in this work, are in agreement with the predictions drawn from the MHD simulations \citep{2010ApJ...720L..26L}. 

\section{Conclusions}
G11 is a filamentary IRDC and it is among the densest types of filaments. Its CO molecular line observations show supersonic turbulence of $\delta v\sim2\,\mathrm{km\,s^{-1}}$. Coordinated with the measurements of magnetic fields in this work, the supersonic turbulence is also sub-Alfv{\'e}nic in the envelope of G11, and increases to trans-Alfv{\'e}nic towards the dense regions of G11. The dynamically important magnetic fields in the envelope of G11 have general implications in the formation of dense filaments. The supersonic and sub-Alfv{\'e}nic turbulence can easily compress gas along the field lines, resulting in density enhanced filament perpendicular to the field lines \citep{2008ApJ...687..354N,2010ApJ...720L..26L}. As soon as the filament becomes supercritical, global collapse starts. Gas in the envelope free-falls onto the supercritical filament. In the case of strong magnetic fields ($50-100\,\mu$G), global collapse is more efficient along the magnetic fields, and this results in higher density structures perpendicular to the strong magnetic fields. \citet{2021MNRAS.503.5425B} found that gravity helps the creation of density structures perpendicular to magnetic fields in their sub-Alfv{\'e}nic models. We suggest that dense filaments like G11 are more easily formed in sub-Alfv{\'e}nic self-gravitating turbulent molecular clouds. Magnetic fields determine the orientation of the filaments formed in such clouds.

\section*{Acknowledgements}
We would like to thank the anonymous referees for their in-depth reviews, which helped us in improving the manuscript. This work is supported by the general grant Nos. U2031202 and 11903083 of the National Natural Science Foundation of China. We are grateful to all the members of the MWISP project, for their long-term support. MWISP is sponsored by the National Key R\&D Program of China with grant No. 2017YFA0402700 and the CAS Key Research Program of Frontier Sciences with grant No. QYZDJ-SSW-SLH047. This paper uses observations made at the South African Astronomical Observatory (SAAO). We thank the SAAO support astronomers and previous observer of the IRSF telescope for their technical support during the observation nights at Sutherland, South Africa, and the SAAO staff members for taking care of our lives at the SAAO. This work has made use of data from the European Space Agency (ESA) mission
{\it Gaia} (\url{https://www.cosmos.esa.int/gaia}), processed by the {\it Gaia} Data Processing and Analysis Consortium (DPAC, \url{https://www.cosmos.esa.int/web/gaia/dpac/consortium}). Funding for the DPAC has been provided by national institutions, in particular the institutions participating in the {\it Gaia} Multilateral Agreement. This work has made use of 353 GHz dust polarisation data, based on observations obtained with Planck (http://www.esa.int/Planck), an ESA science mission with instruments and contributions directly funded by ESA Member States, NASA, and Canada. This work has made use of NASA’s Astrophysics Data System. This research is based on data products from observations made with ESO Telescopes at the La Silla or Paranal Observatories under ESO programme ID 179.B-2002, and has made use of the services of the ESO Science Archive Facility.

We use {\it Python} packages NUMPY \citep{2020Natur.585.7825H}, SCIPY \citep{2020NatMe..17..261V}, ASTROPY \citep{2013A&A...558A..33A,2018AJ....156..123A}, and Spectral-cube \citep{2015ASPC..499..363G} to analyse the data used in this work. We use SAOImage DS9 \citep{2003ASPC..295..489J,2019zndo...2530958J} and {\it Python} pacakge MATPLOTLIB \citep{2007CSE.....9...90H} for data visualization. 

\section*{Data Availability}
The full version of Table~\ref{tbl:Kpol} and MWISP CO data of G11 are available in https://www.scidb.cn/en, at https://dx.doi.org/10.57760/sciencedb.01942 \citep{G11_table}.



\bibliographystyle{mnras}
\bibliography{myrefs} 




\appendix

\section{$^{13}$CO $J=$1-0 moment 0 and moment 2 maps of the G11 filament}\label{Sect:appendix}
\begin{figure*} 
   \centering
   \includegraphics[width=0.95\textwidth]{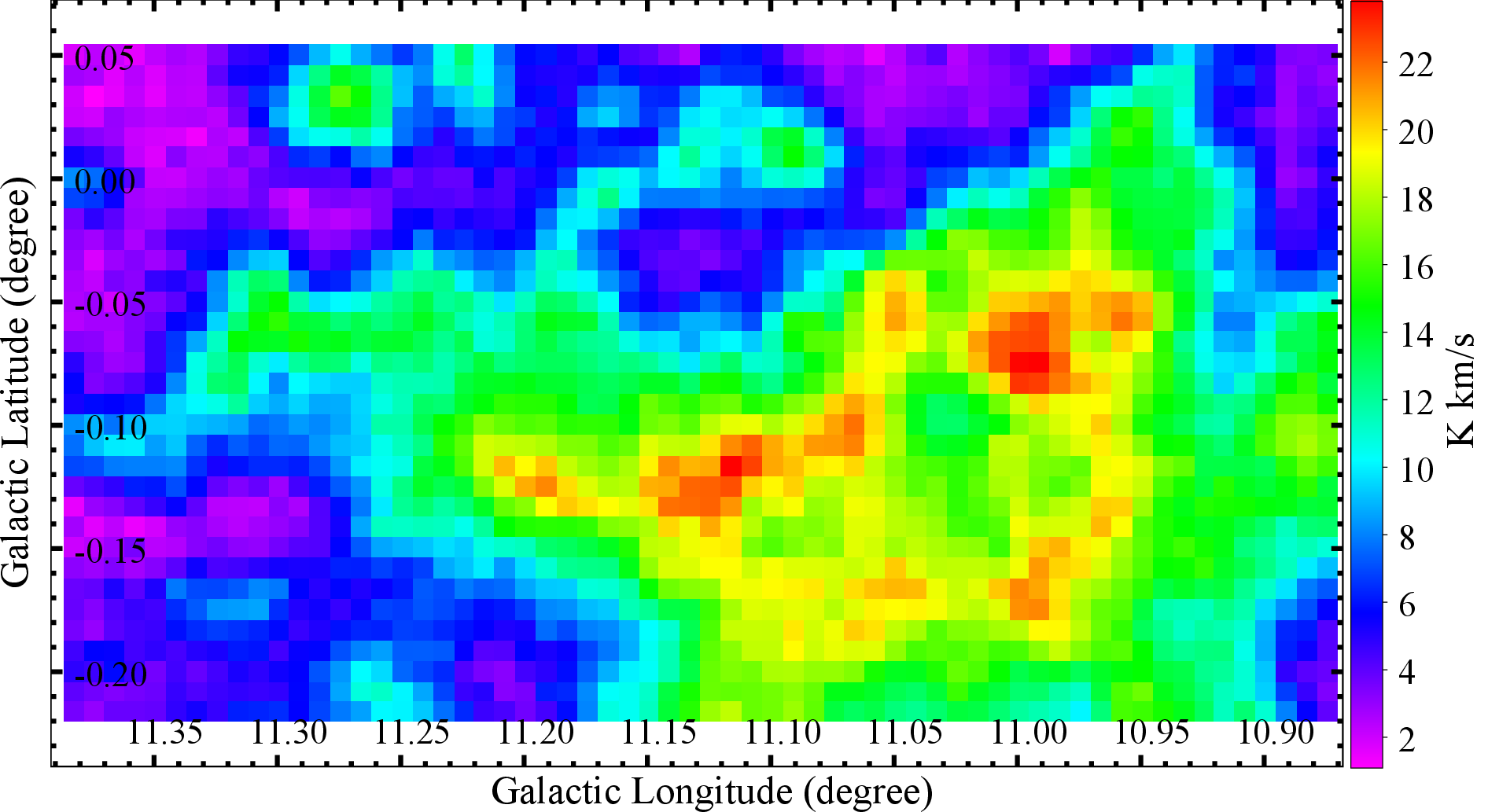}
   \caption{Moment 0 map of the $^{13}$CO $J$=1-0 emission in the velocity range of $25-37\,\mathrm{km\,s^{-1}}$.}
   \label{Fig:mom0}
 \end{figure*}

\begin{figure*} 
   \centering
   \includegraphics[width=0.95\textwidth]{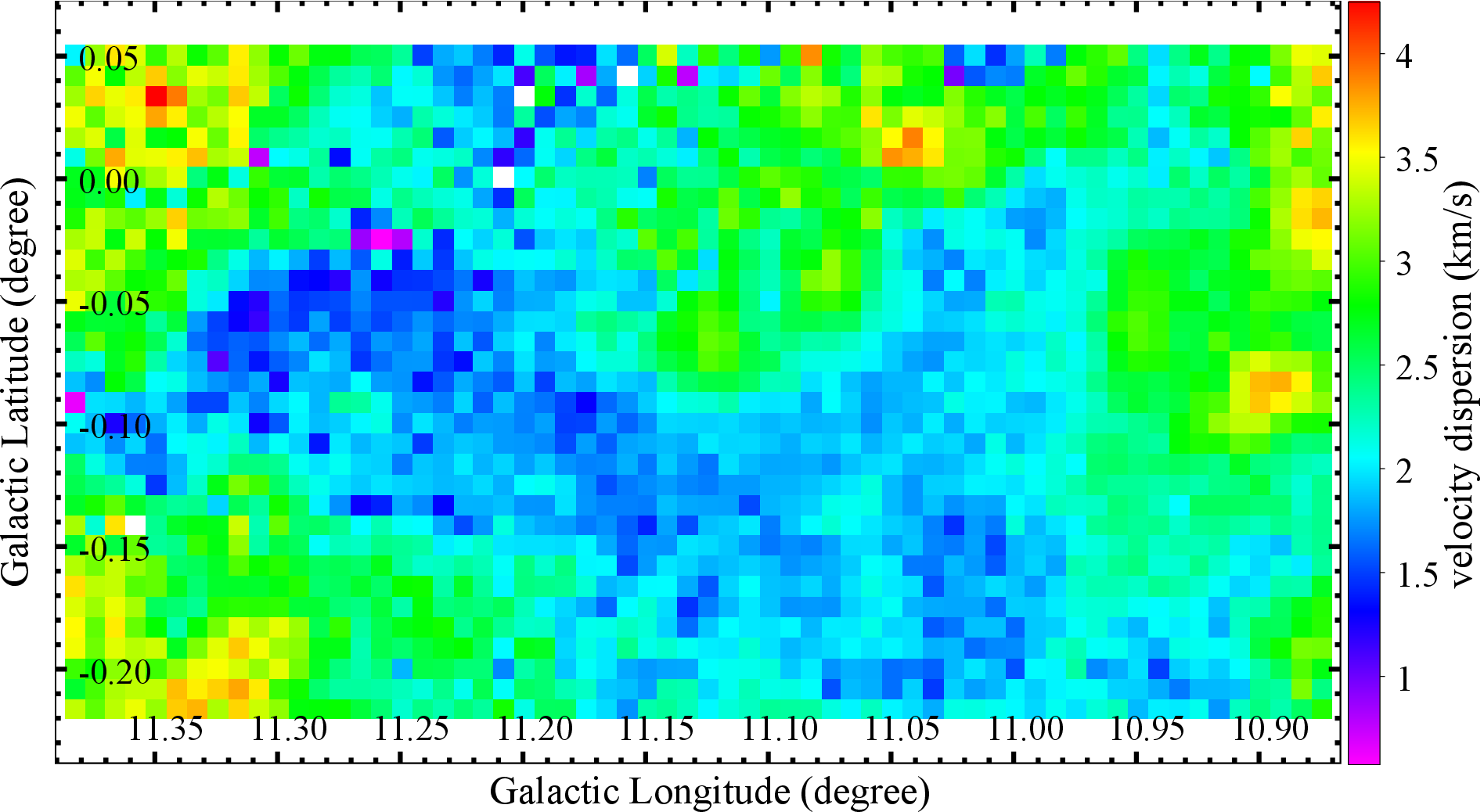}
   \caption{Moment 2 map of the $^{13}$CO $J$=1-0 emission in the velocity range of $25-37\,\mathrm{km\,s^{-1}}$.}
   \label{Fig:mom2}
 \end{figure*}

The moment 0 and moment 2 maps of the $^{13}$CO $J$=1-0 emission in the velocity range of $25-37\,\mathrm{km\,s^{-1}}$ are shown in Figures~\ref{Fig:mom0} and \ref{Fig:mom2}, respectively. 

\bsp	
\label{lastpage}
\end{CJK*}
\end{document}